\documentclass[twocolumn,apj]{openjournal}\usepackage{graphicx}\usepackage{xcolor}\usepackage{natbib}
\usepackage{verbatim}
\usepackage{comment} 
\usepackage{amsmath}
\usepackage{calrsfs}
\usepackage{rotating}   
\usepackage{booktabs}   
\DeclareMathAlphabet{\pazocal}{OMS}{zplm}{m}{n}

\makeatletter
\def\@hex@@Hex#1%
 {\if a#1A\else \if b#1B\else \if c#1C\else \if d#1D\else
  \if e#1E\else \if f#1F\else #1\fi\fi\fi\fi\fi\fi \@hex@Hex}
\makeatother
\definecolor{afcolor}{HTML}{b3443c}

\def\Msun{M_\odot}
\def\HH{$\rm H_{2}$~}
\def\msun{\Msun}

\revised{\today}

\graphicspath{{./}{figures/}}

\usepackage{xurl}
\usepackage[
    colorlinks=true,
    linkcolor=blue,
    citecolor=blue,
    urlcolor=blue
]{hyperref}

\hypersetup{
    pdfauthor={Andrea Ferrara},
    pdftitle={No Blue without Red}
}
\usepackage[utf8]{inputenc}
\usepackage[T1]{fontenc}

\DeclareUnicodeCharacter{2243}{\ensuremath{\simeq}}
\DeclareUnicodeCharacter{2500}{--}
\DeclareUnicodeCharacter{224D}{\ensuremath{\asymp}}
\DeclareUnicodeCharacter{223C}{\ensuremath{\sim}}
\DeclareUnicodeCharacter{03B1}{\ensuremath{\alpha}}

\begin{document}



\def\be{\begin{equation}}
\def\ee{\end{equation}}
\newcommand\code[1]{\textsc{\MakeLowercase{#1}}}
\newcommand\quotesingle[1]{`{#1}'}
\newcommand\quotes[1]{``{#1}"}
\def\gsim{\lower.5ex\hbox{\gtsima}} 
\def\lsim{\lower.5ex\hbox{\ltsima}} 
\def\gtsima{$\; \buildrel > \over \sim \;$} 
\def\ltsima{$\; \buildrel < \over \sim \;$} \def\gsim{\lower.5ex\hbox{\gtsima}} 
\def\lsim{\lower.5ex\hbox{\ltsima}} 
\def\simgt{\lower.5ex\hbox{\gtsima}} 
\def\simlt{\lower.5ex\hbox{\ltsima}}

\def\msun{{\rm M}_{\odot}}
\def\lsun{{\rm L}_{\odot}}
\def\dsun{{\cal D}_{\odot}}
\def\fsun{\xi_{\odot}}
\def\zsun{{\rm Z}_{\odot}}
\def\msunyr{\msun {\rm yr}^{-1}}
\def\gdens{\msun\,{\rm kpc}^{-2}}
\def\sfrdens{\msun\,{\rm yr}^{-1}\,{\rm kpc}^{-2}}

\def\mum{\mu {\rm m}}
\newcommand{\angstrom}{\mbox{\normalfont\AA}}
\def\cc{\rm cm^{-3}}
\def\uflux{{\rm erg}\,{\rm s}^{-1} {\rm cm}^{-2} }

\def\fdust{\xi_{d}}
\def\fesc{f_{\rm esc}}
\def\td{\tau_{sd}}
\def\Sg{$\Sigma_{g}$}
\def\S*{$\Sigma_{\rm SFR}$}
\def\Ssfr{\Sigma_{\rm SFR}}
\def\Sgas{\Sigma_{\rm g}}
\def\Sstar{\Sigma_{\rm *}}
\def\Sesc{\Sigma_{\rm esc}}
\def\Srad{\Sigma_{\rm rad}}
\def\sSFR{\rm sSFR}

\def\Dsolar{${\cal D}/\dsun$}
\def\Zsolar{$Z/\zsun$}
\def\DDsolar{\left( {{\cal D}\over \dsun} \right)}
\def\ZZsolar{\left( {Z \over \zsun} \right)}
\def\kms{{\rm km\,s}^{-1}\,}
\def\skms{$\sigma_{\rm kms}\,$}

\def\Scii{$\Sigma_{\rm [CII]}$}
\def\Sciimax{$\Sigma_{\rm [CII]}^{\rm max}$}
\def\CII{\hbox{[C~$\scriptstyle\rm II $]~}}
\def\CIII{\hbox{C~$\scriptstyle\rm III $]~}}
\def\OII{\hbox{[O~$\scriptstyle\rm II $]~}}
\def\OIII{\hbox{[O~$\scriptstyle\rm III $]~}}
\def\HH{\hbox{H$_2$}~} 
\def\HI{\hbox{H~$\scriptstyle\rm I\ $}} 
\def\HII{\hbox{H~$\scriptstyle\rm II\ $}} 
\def\CIion{\hbox{C~$\scriptstyle\rm I $~}}
\def\CIIion{\hbox{C~$\scriptstyle\rm II $~}}
\def\CIIIion{\hbox{C~$\scriptstyle\rm III $~}}
\def\CIVion{\hbox{C~$\scriptstyle\rm IV $~}}
\def\nhh{n_{\rm H2}}
\def\nhi{n_{\rm HI}}
\def\nhii{n_{\rm HII}}
\def\fhh{x_{\rm H2}}
\def\fhi{x_{\rm HI}}
\def\fhii{x_{\rm HII}}
\def\fd{f^*_{\rm diss}} 
\def\ks{\kappa_{\rm s}}

\def\cyan{\color{cyan}}
\definecolor{apcolor}{HTML}{b3003b}
\definecolor{afcolor}{HTML}{800080}
\definecolor{lvcolor}{HTML}{DF7401}
\definecolor{mdcolor}{HTML}{01abdf} 
\definecolor{cbcolor}{HTML}{ff0000}
\definecolor{sccolor}{HTML}{cc5500} 
\definecolor{sgcolor}{HTML}{00cc7a}
{Missing definitions.tex; supply original definitions file}}

\title{No Blue without Red: Evolutionary Properties of Super-Early Galaxies}

\author{A. Ferrara$^{1}$, G. Rodighiero$^{2}$, S. Carniani$^{1}$,
Z. Zhang$^{1,3,4}$, M. Kohandel$^{1}$, B. Das$^{1}$\\[1ex]
\small $^{1}$Scuola Normale Superiore, Piazza dei Cavalieri 7, 50126 Pisa, Italy\\
\small $^{2}$Dipartimento di Fisica e Astronomia ``G. Galilei'', Universit\`a di Padova, Vicolo dell'Osservatorio 3, 35131 Padova, Italy\\
\small $^{3}$National Astronomical Observatories, Chinese Academy of Sciences, 20A Datun Road, Chaoyang District, Beijing 100101, P. R. China\\
\small $^{4}$School of Astronomy and Space Science, University of Chinese Academy of Sciences, Beijing 100049, P. R. China\\
\small Corresponding author: andrea.ferrara@sns.it}

\begin{abstract}
The discovery of numerous luminous, super-early galaxies at $z>10$ by {\it JWST} has revealed a striking diversity in their ultraviolet (UV) properties, ranging from extremely blue, dust-poor systems to a smaller population of significantly reddened sources. We investigate the physical origin of this diversity within the framework of the Attenuation-Free Model (AFM), in which radiation-driven outflows redistribute dust to large galactic radii, reducing the effective attenuation. {Starting from a sample of 32 spectroscopically confirmed super-early
galaxies, we apply the detailed AFM analysis to a subset of 17 systems
with suitable size constraints} and derive their key physical properties, including halo mass, star formation efficiency, metallicity, and outflow extent. We find that these systems reside in massive halos ($\log M/M_\odot \sim 10.7$) and exhibit moderate ($0.01 \lesssim \epsilon_* \lesssim 0.05$) star formation efficiencies, while frequently reaching super-Eddington conditions that trigger powerful outflows. {Within the AFM framework, we propose a possible evolutionary sequence}
in which galaxies transition from a dust-obscured ``Red Monster'' phase
to a UV-bright ``Blue Monster'' phase as outflows clear their central
regions. The recently confirmed red galaxy EGS-z11-R0 at $z=11.45$ is naturally interpreted as a system observed during this obscured phase. Compact ($r_e \lesssim 150$ pc) sources are instead difficult to reconcile within AFM; we speculate that their emission is dominated by an AGN. Our results provide a {coherent} interpretation of super-early galaxy properties and highlight the key role of radiation-driven outflows in shaping galaxy evolution at cosmic dawn. Future observations with {\it JWST} and ALMA will be essential to test these predictions and further constrain the nature of the earliest galaxies.
\end{abstract}

\section{Introduction} \label{sec:intro}
The first few hundred million years after the Big Bang represent a critical phase in galaxy formation, when the earliest stellar systems assembled, enriched their interstellar media with heavy elements, and contributed to cosmic reionization (for reviews see \citealt{sommerville:2015, Dayal18}). Since its commissioning, the James Webb Space Telescope (JWST) has revolutionized our view of this epoch, revealing a surprisingly abundant population of luminous galaxies at \(z > 10\), corresponding to cosmic ages \(\lesssim 500\) Myr (e.g., \citealt{Naidu22, Harikane22, Castellano23, Bunker23, Finkelstein24, Curtis23, Carniani24a,  Naidu25}). A defining characteristic of the majority of these systems is their extremely blue rest-frame ultraviolet (UV) continuum slopes, often reaching \(\beta_{\mathrm{UV}}\lesssim -2.5\) \citep[]{Morales_2024, Cullen24, Saxena24, Austin24, Dottorini24}. Such steep slopes are typically interpreted as evidence for young, metal-poor stellar populations with minimal dust attenuation, earning them the nickname ``Blue Monsters'' \citep[]{Ferrara23, Ziparo23}.

The properties of Blue Monsters -- including their bright UV luminosities (\(M_{\mathrm{UV}} \lesssim -20\)), compact sizes (\(r_e \approx 200\) pc), and large stellar masses (\(M_\star \approx 10^9\,M_\odot\)) -- have posed significant challenges to theoretical models based on pre-JWST expectations. In response, the so-called ``attenuation-free model'' (AFM) has been proposed \citep[]{Ferrara23, Ziparo23, Fiore23, Ferrara24a, Ferrara24b}. According to AFM, the extreme UV brightness and blue colors of these galaxies result from extremely low dust attenuation conditions, rather than from an intrinsic lack of dust. In this scenario, dust is produced by stars in standard net amounts but is subsequently pushed to kiloparsec scales by radiation-driven outflows, dramatically reducing the UV attenuation for a fixed dust mass (since \(A_V \propto r^{-2}\)). The AFM has successfully explained the shape of the UV luminosity functions at \(z>10\), the evolution of the cosmic star-formation rate density, and the star-formation histories of individual galaxies \citep{Ferrara24a, Ferrara24b, Ferrara25a}.

A key prediction of the AFM is that the dust-to-stellar mass ratio \(\xi_d\) in Blue Monsters should be comparable to that measured in somewhat lower-redshift (\(z\sim7\)) galaxies, i.e., \(\log\xi_d \sim -2\) \citep{Ferrara22a, Dayal22, Sommovigo22}. However, direct observational constraints from JWST/NIRSpec spectra of 15 spectroscopically confirmed \(z>10\) galaxies instead indicate \(\log\xi_d \lesssim -4\) \citep[]{Ferrara25b}. This striking discrepancy suggests that standard dust production and destruction physics cannot reproduce the observed UV transparency of Blue Monsters. Within the AFM framework, this tension is resolved if dust is efficiently displaced from the central regions by radiation-driven outflows: while the total dust mass remains high, its redistribution to large radii significantly reduces the effective attenuation along the line of sight, yielding the observed UV transparency.

While super-early galaxy simulations show the presence of strong outflows (see e.g. \citealt{Kohandel25, Ceverino24}), direct observational evidence for such outflows in \(z>10\) galaxies has recently begun to emerge. Notably, deep JWST/NIRSpec observations of GS-z14-0 at \(z=14.18\), have revealed spatially offset CIII]$\lambda\lambda$1907,1909 emission extending up to \(\sim400\) pc from the stellar continuum, with a luminosity three times brighter than that measured in the nuclear spectrum \citep[]{Carniani26}. The lack of a corresponding feature in ultra-deep NIRCam medium-band imaging implies that the emitting gas is diffuse rather than point-like, consistent with a scenario in which carbon-enriched gas has been expelled from the galaxy by ongoing or past outflows. From the CIII] luminosity, the authors  derive a mass outflow rate of \(\dot{M}_{\mathrm{out}} \approx 55 M_{\odot}\,\mathrm{yr}^{-1}\), corresponding to a mass-loading factor of \(\eta = \dot{M}_{\mathrm{out}} / \mathrm{SFR} \approx 1-5.5\). This suggests that outflows are highly efficient at early times, expelling gas at rates comparable to or exceeding then SFR. 

Further support to the outflows scenario is provided by \citet[]{Tripodi26} who presented a spatially resolved stacked analysis of 287 LAEs at  $z>4$  observed with JWST/NIRSpec prism spectroscopy. Similarly, \citet[]{Maiolino24b} in a study dedicate to GN-z11 ($z=10.6$) find a deep, blue-shifted CIV$\lambda$1549 absorption trough, tracing an outflow with a velocity of $\sim 800\ \kms$. The outflow is likely driven by radiation pressure from the combined stellar + AGN luminosity. 

Further indirect evidence comes from the galaxy UNCOVER-37126 at \(z \approx 10.25\). \citet{Marques26} find that this galaxy is undergoing an “ISM-naked” starburst phase, in which a strong outflow has cleared the remaining gas from its stellar core, allowing most LyC photons to escape. 

Together, these observations provide support for the AFM framework, demonstrating that radiatively driven outflows are already operating at the earliest cosmic epochs, efficiently clearing dust and gas from the central regions of galaxies and thereby enabling the emergence of the bright, blue systems that JWST has revealed.

While the AFM provides a compelling framework for the dominant population of blue, dust-poor galaxies at \(z>10\), recent studies have uncovered a small but intriguing population of candidate \citep[]{Rodighiero23, Mitsuhashi25, Tang25} or spectroscopically confirmed \citep[]{Kokorev25, Napolitano25, Donnan26, Rodighiero26} high-redshift galaxies exhibiting significantly redder UV continua, i.e., \(\beta_{\mathrm{UV}} > -1.5\) . These objects deviate markedly from the canonical \(M_{\mathrm{UV}}\)–\(\beta_{\mathrm{UV}}\) relation established at \(z\sim6\)–\(9\) and extended to \(z\sim10\)–\(12\) with JWST. Their red slopes are difficult to reconcile with simple models of chemically primitive galaxies given the short age of the Universe at these epochs (\(\lesssim400\) Myr). Instead, significant dust attenuation (\(A_V \sim 0.5\)–\(1\) mag, \citealt{Kohandel25}) or extreme nebular continuum emission from dense, highly ionized gas have been proposed as possible explanations \citep{Mitsuhashi25, Katz25}.

A landmark in this context is the recent spectroscopic confirmation of EGS-z11-R0 at \(z = 11.452\pm0.021\) \citet[]{Rodighiero26}, the most distant red galaxy identified and confirmed to date. Discovered serendipitously through inspection of public JWST/NIRSpec data in the CEERS field, EGS-z11-R0 exhibits a remarkably flat UV continuum slope \(\beta_{\mathrm{UV}} \sim -1.0\), placing it well above the canonical \(M_{\mathrm{UV}}\)–\(\beta_{\mathrm{UV}}\) relation at \(z\sim10\)–\(12\). The source shows significant detections of high-ionization UV emission lines such as CIV and CIII], and its spectral energy distribution (SED) requires substantial dust attenuation (\(A_V\sim {1}\) mag) and a stellar mass \(\log(M_\star/M_\odot)\sim 8.5\)–\(9.5\). While a moderate AGN contribution cannot be excluded, the data robustly indicate a dust-enriched system observed during an early, possibly obscured phase of galaxy assembly.

The existence of such dust-reddened systems at \(z>10\) has important implications for our understanding of early chemical enrichment and galaxy evolution. Several theoretical models have explored the rapid buildup of galaxies at cosmic dawn. 

\citet[]{Ferrara24a} proposed that galaxies at \(z>10\) initially undergo a radiative feedback-regulated phase leading to bursty star formation. When dust shielding becomes important, the galaxy switches to a smooth, substantially obscured star formation regime—dubbed the ``Red Monster'' phase. During this obscured phase, which lasts \(\simeq20\%\) of the galaxy lifetime, \(\simeq70\%\) of the observed stars are formed. 

Eventually, as the galaxy becomes super-Eddington, a powerful radiation-driven outflow clears most of the dust, and the galaxy transitions into a ``Blue Monster'' dominating the bright end of the UV luminosity function. Other theoretical frameworks, such as the feedback-free model \citep[]{Dekel23, Li23} or density-regulated star formation in semi-analytic models \citep[]{Somerville25}, have also emphasized the possibility of rapid, early stellar mass assembly.

{An alternative class of explanations invokes stochastic or bursty star
formation. In this picture, substantial short-timescale fluctuations in the
SFR produce a broad distribution of UV luminosities at fixed halo mass,
allowing intrinsically more numerous, lower-mass systems to be temporarily
up-scattered into the bright end of the UV luminosity function
\citep[e.g.][]{Pallottini23, Sun23, Mason23, Gelli25}. Such models can therefore
alleviate the apparent excess of luminous $z>10$ galaxies without requiring
them to reside exclusively in extremely rare halos. }

{More generally, the
observed UV properties may also be affected by nebular continuum emission,
variations of the stellar IMF, feedback-free star formation, and AGN
activity. These possibilities are physically distinct from the AFM
interpretation considered here and imply that the latter should not be
regarded as a unique explanation of the current data.}

Observational evidence increasingly points toward rapid early stellar assembly at these epochs. \citet[]{Santini25} showed that several \(z>10\) systems exhibit unexpectedly high stellar masses and specific star-formation rates, indicating that substantial stellar assembly had already occurred within the first few hundred million years. 

However, robust spectroscopic confirmation of strongly dust-reddened galaxies at \(z>10\) has until now been lacking. Most red candidates identified so far rely on broadband photometry, where degeneracies between dust, age, nebular emission, and AGN activity complicate the interpretation. The spectroscopic confirmation of EGS-z11-R0 thus provides a critical anchor for understanding whether chemically evolved, dust-enriched galaxies were already in place within the first \(\sim400\) Myr of cosmic history.

In this work, we investigate the origin of the observed diversity of super-early galaxies at $z>10$ within the framework of the AFM. We apply the model to a sample of spectroscopically confirmed galaxies to derive their key physical properties -- including halo mass, star formation efficiency, metallicity, and outflow extent -- and to directly test the role of radiation-driven outflows in shaping their observable characteristics. 

{We show that the coexistence of extremely blue and significantly reddened
systems can be consistently interpreted as different stages of a possible
common evolutionary sequence}, in which galaxies transition from a dust-obscured phase to a UV-bright one as outflows efficiently redistribute gas and dust. 

Within this framework, recently identified red systems, such as EGS-z11-R0, are interpreted as galaxies caught during an early, obscured phase of their evolution. This approach provides a physically motivated picture for the buildup of dust, metals, and stellar mass at cosmic dawn, and establishes a direct link between the observed properties of $z>10$ galaxies and their underlying evolutionary pathways.

This paper is structured as follows. In Sec. \ref{sec:AFM} we outline the AFM and its key physical ingredients.  In Sec. \ref{sec:results} we apply the model to a sample of spectroscopically confirmed $z>10$ galaxies and derive their physical properties. We also interpret the results in terms of a {possible} evolutionary scenario linking dust-obscured and UV-bright phases. In Sec. \ref{sec:compact} we discuss the nature of the compact galaxy subset and the possible role of AGN. Finally, Sec. \ref{sec:summary} summarizes our conclusions. We adopt a $\Lambda$CDM cosmological model\footnote{Throughout the paper, we assume a flat Universe with the following cosmological parameters: $\Omega_m = 0.3075$, $\Omega_{\Lambda} = 1- \Omega_{\rm M}$, and $\Omega_{b} = 0.0486$,  $h=0.6774$, $\sigma_8=0.826$, where $\Omega_{m}$, $\Omega_{\Lambda}$, and $\Omega_{b}$ are the total matter, vacuum, and baryon densities, in units of the critical density; $h$ is the Hubble constant in units of $100\,\kms \rm Mpc^{-1}$, and $\sigma_8$ is the late-time fluctuation amplitude parameter \citep{planck:2015}.}, and a Salpeter $1-100\ M_\odot$ initial mass function.


\section{Attenuation-Free Model} \label{sec:AFM}
The Attenuation-Free Model (AFM, \citealt{Ferrara23}) provides a comprehensive framework to interpret the properties of super-early ($z>10$) galaxies, and test and explore their implications. In addition to reproduce the observed  UV luminosity functions and evolution of the cosmic star formation rate density \citep[]{Ferrara24a}, AFM also makes testable predictions on the galaxy dust content \citep[]{Ferrara25b}, sub-mm observability \citep[]{Ferrara25a}, escape fraction of Lyman continuum \citep[]{Ferrara25c} and Ly$\alpha$ \citep[]{Ferrara24a} photons. Furthermore, it has been used to successfully derive the star formation history of super-early galaxies, like e.g. GS-z14-0 \citep[]{Ferrara24b}. 

With the aim of understanding the origin of super-early galaxies diversity, and in particular their brightness and spectral colors, we apply AFM to the spectroscopically confirmed sources at $z>10$. In the following we recall the AFM main features and key equations.  

\subsection{Galaxy sizes} \label{subsec:sizes}
We start by deriving the half-light (or effective) galaxy radius, $r_e$, under the standard assumption that the specific angular momentum of the halo is conserved during disk formation \citep[]{Mo10}. Hence, $r_e$ is related to the halo virial radius, $r_{\rm vir}$, via the spin parameter, $\lambda$, as 
\be
r_e = 1.678\frac{\lambda}{\sqrt2} r_{\rm vir} = 0.1678 \frac{\lambda}{\sqrt2} \frac{v_c}{H(z)},
\label{eq:re}
\ee
where $v_c$ is the halo circular velocity.
The numerical pre-factor comes from the implicit assumption of an exponential disk, for which the effective radius is $r_e=1.678$ times the disk scale-length. As shown by numerical simulations \citep{Angel16},  $\lambda$ depends very weakly on halo mass $M$ and redshift. Its distribution is approximately log-normal with mean value $\lambda=0.035$ and $\sigma_{\log\lambda}\approx 0.22$ dex scatter \citep[]{Mo10,Semenov26}. 

\begin{figure}
\centering\includegraphics[width = 0.48 \textwidth]{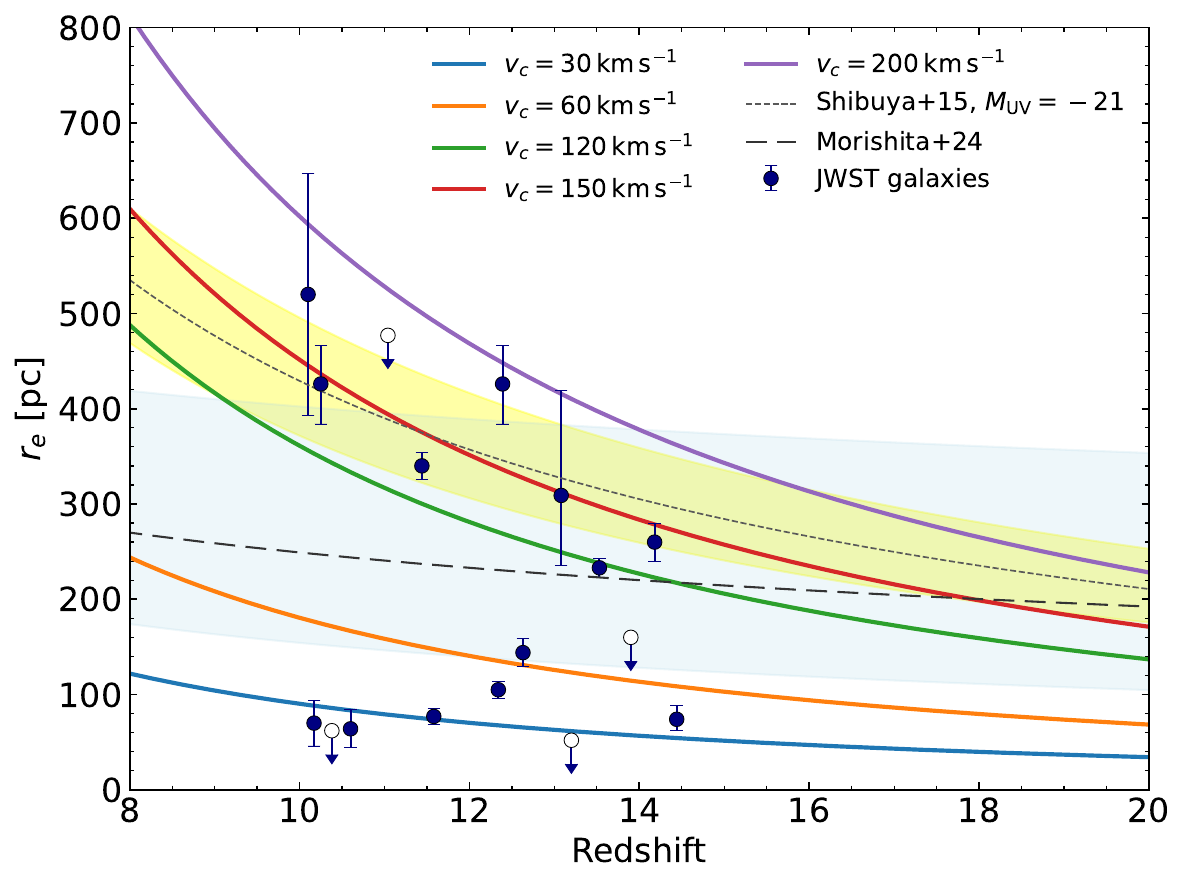}
\caption{Predicted redshift evolution of the effective galaxy radius, $r_e$ (eq. \ref{eq:re}), for different values of the halo circular velocities, $v_c$, as shown by the legend. The data points represent our super-early galaxy sample presented in Tab. \ref{tab:observed_props} along with the references. Open points indicate upper limits. Also shown are the best-fit curves and $1\sigma$ errors (shaded region) to high-$z$ galaxy data from \citet[]{Shibuya15} and \citet[]{Morishita_2024}.  
} 
\label{fig:sizes}
\end{figure}

In Fig. \ref{fig:sizes} we compare $r_e(v_c)$ from eq. \ref{eq:re} with the observed effective radii of the spectroscopically confirmed super-early galaxies taken from the literature. Also shown is the extrapolation of the size evolution obtained by \cite{Shibuya15} for bright ($M_{\rm UV} \simeq -21$) Lyman Break Galaxies, and the same relation obtained by \citet[]{Morishita_2024} for 341 galaxies at $5 < z < 14$, with 109 having spectroscopic redshift measurements from the literature. 

%
%
%
%
%
\begin{figure*}
\centering\includegraphics[width = 0.49 \textwidth]{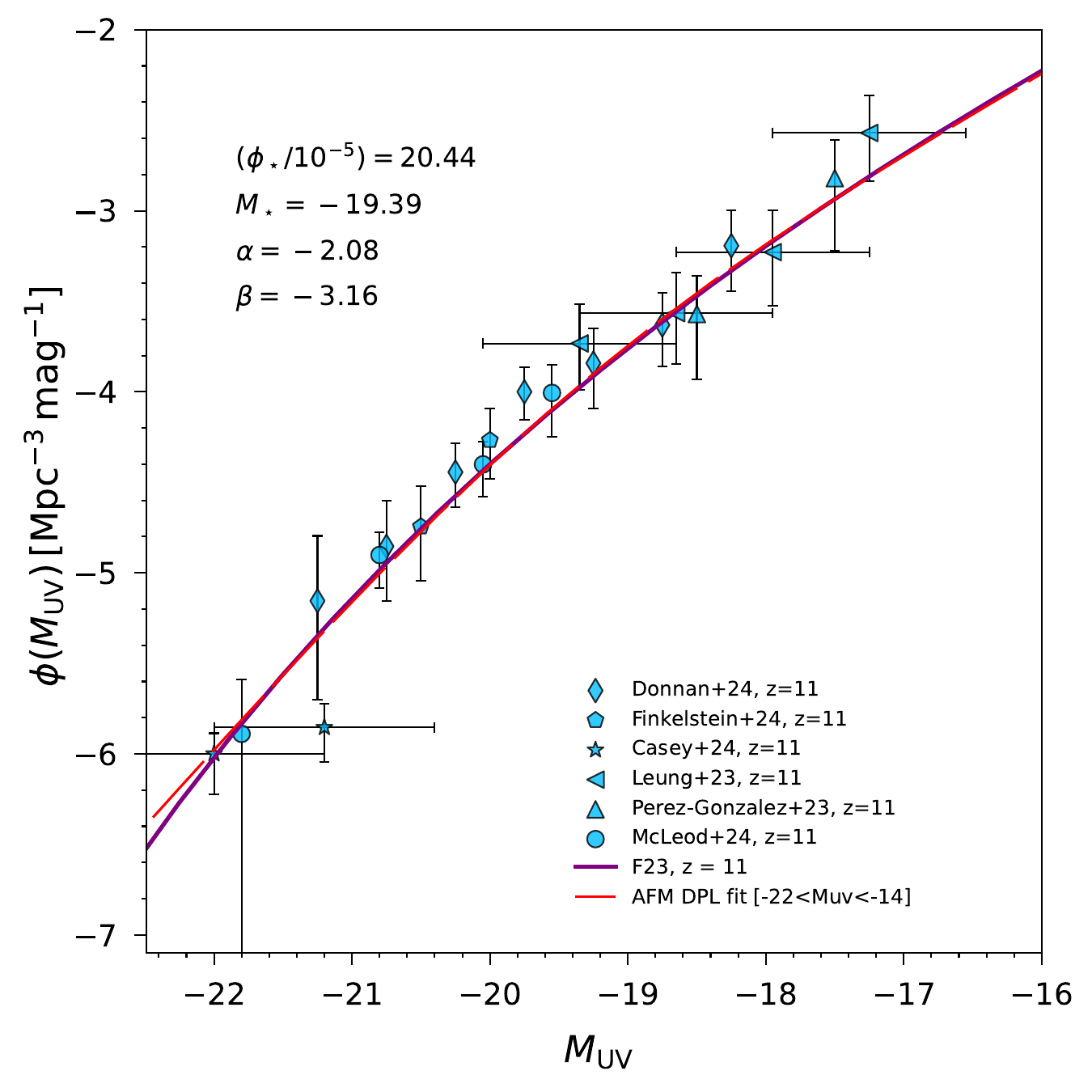}
\centering\includegraphics[width = 0.49 \textwidth]{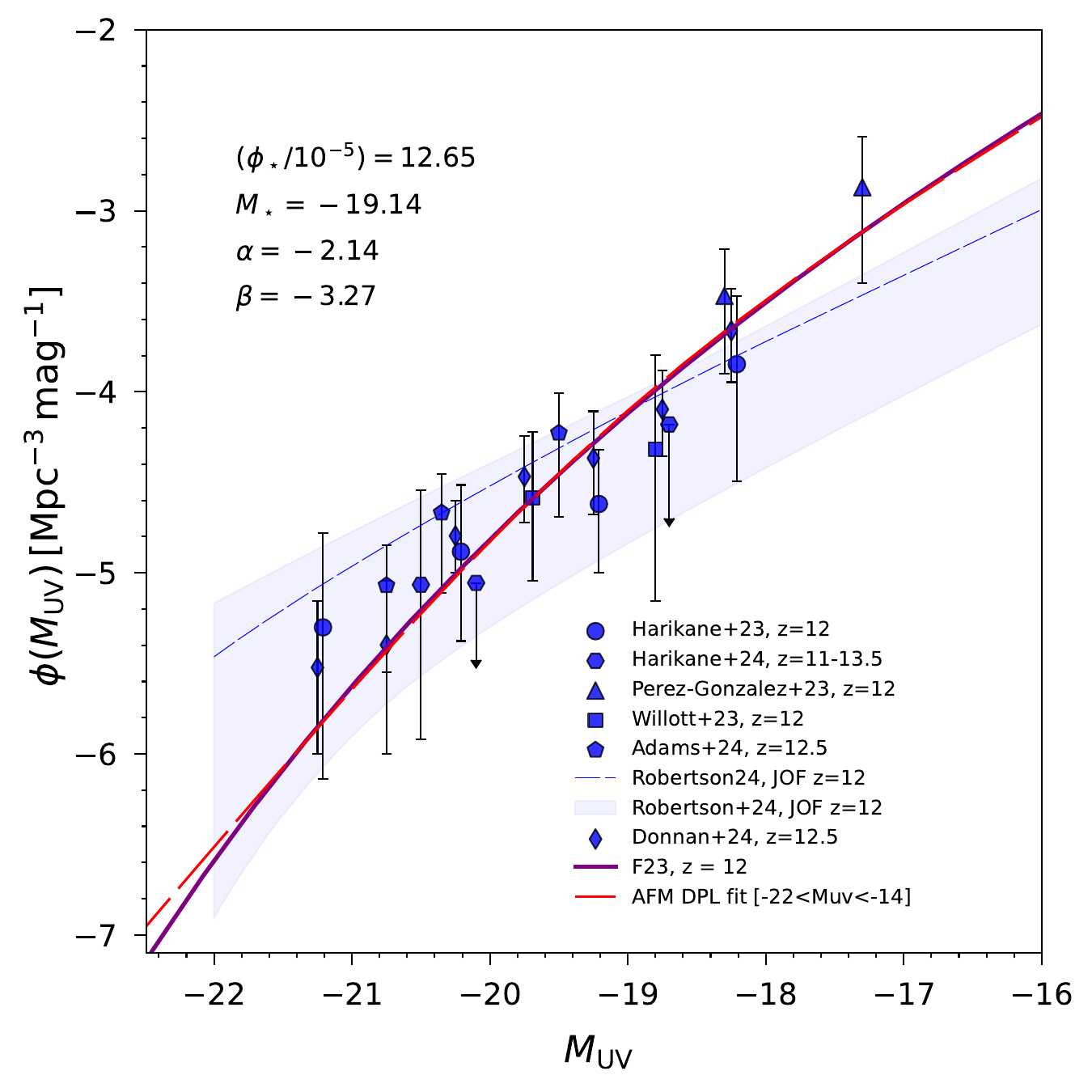}
\caption{Attenuation-Free Model (AFM) predictions for the UV luminosity function at $z\simeq 11$ (left panel) and $z\simeq 12$ (right) compared to available JWST data. The model is well fitted by a double-power law (red curve) of the form $\phi(M_{\rm UV})= \phi_* [10^{(1+\alpha)x} + 10^{(1+\beta)x}]^{-1}$, where $x= -0.4(M_*-M_{\rm UV})$, with parameters shown in the panels.   Data points are taken from 
\citet[]{Donnan24, Finkelstein23, Casey23, Leung23, Perez23, McLeod23, Harikane23c, Harikane24, Willott24, Adams24, Robertson24}
} 
\label{fig:AFM_UVLF}
\end{figure*}

As already noted by \citet[]{Harikane25}, super-early galaxies seem to fall in two distinct categories featuring either extended or compact morphologies. The first group is consistent with both the \citet[]{Shibuya15} extrapolation and with our model for $v_c\sim 200\ \kms$ galaxies, showing $r_e$ values $\simgt 250$ pc. In the range $10 < z < 14$ such circular velocity corresponds to halo masses of $\simeq 10^{11} M_\odot$. The second group includes galaxies with $r_e \simlt 150$ pc, or $v_c < 80\ \kms$. We note that the \citet[]{Morishita_2024} best-fit seems to run in between the two distributions.  

The origin of this dual distribution is unclear. The sharp separation is hard to be reconciled with the smooth, log-normal distribution of spin parameters, or smaller ($\simlt 8\times 10^9\ M_\odot$) halo masses given their generally comparable inferred stellar masses ($\approx 10^{8-9}\ M_\odot$, see Tab. \ref{tab:observed_props}). As we discuss in Sec. \ref{sec:compact}, the most plausible scenario for these compact galaxies is that their UV light is dominated by AGN activity, as indeed suggested for GN-z11, GHZ2 and GHZ9, based on the detection of high-ionization emission lines \citep[]{Maiolino24b, Castellano24, Naidu25, Napolitano25}.

%
%
\begin{figure}
\centering\includegraphics[width = 0.46 \textwidth]{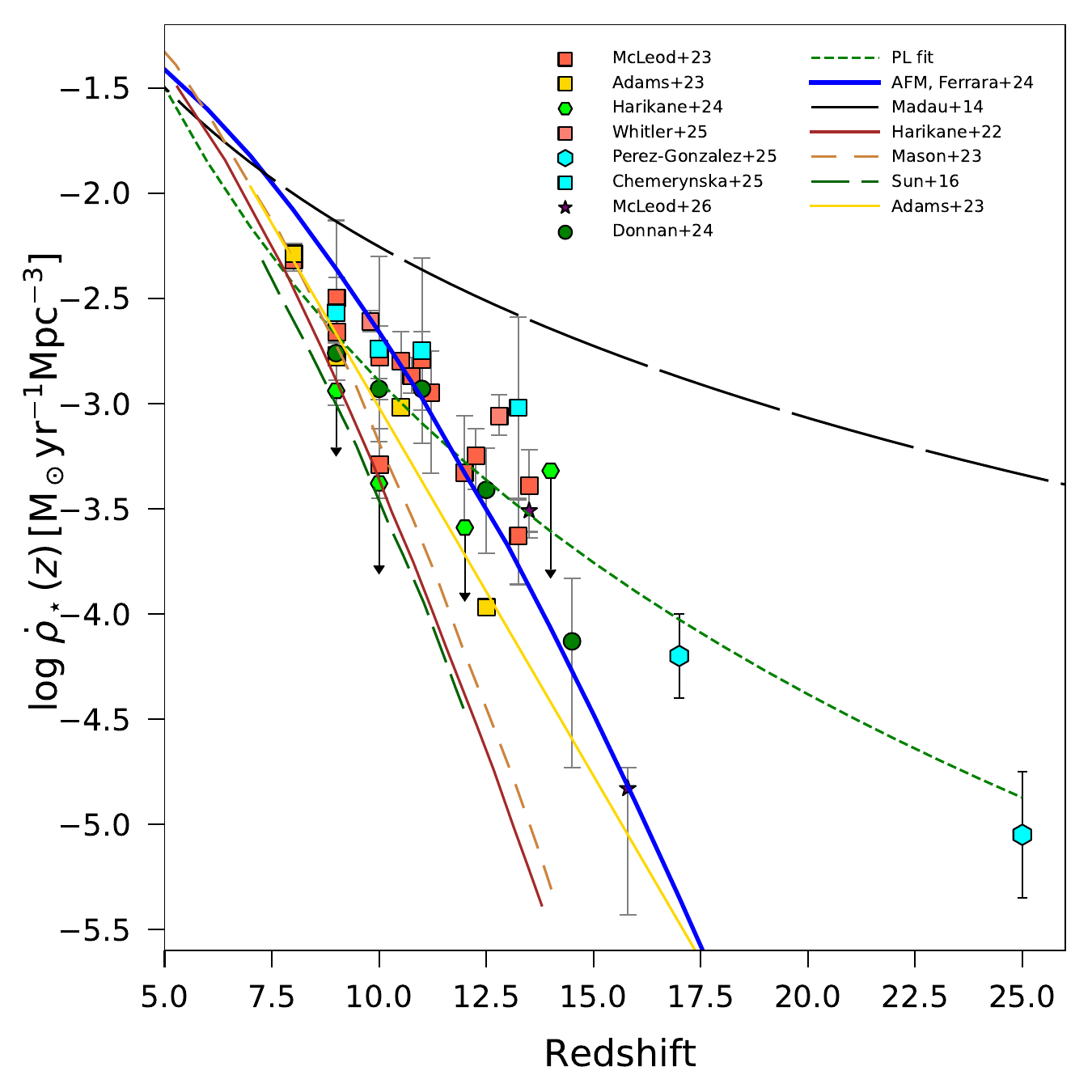}
\caption{The comoving star formation rate density evolution predicted by AFM (blue line) is compared with available data and other models (lines) in the literature. 
Data are taken from \citet[]{McLeod23, Adams24, Harikane24, Whitler23, Perez25, Chemerynska24, Madau14, Harikane22, Mason23, Sun16, Adams23, Donnan24, McLeod26}. Also shown is a power-law fit (green line) to the data.
} 
\label{fig:AFM_SFRD}
\end{figure}
%

%
%
\begin{table*}
\centering
\small
\caption{Observed properties of spectroscopically confirmed galaxies at $z>10$. References are given in the legend.}
\label{tab:observed_props}   
\begin{tabular}{lrrrrrrrrr}
\hline\hline
\multicolumn{1}{l}{Name} & \multicolumn{1}{c}{$z$} & \multicolumn{1}{c}{$M_{\rm UV}$} & \multicolumn{1}{c}{$\beta$} & \multicolumn{1}{c}{SFR} [$M_\odot/\rm yr$] & \multicolumn{1}{c}{$\log (M_\star/M_\odot)$} & \multicolumn{1}{c}{$r_e$ [pc]} & \multicolumn{1}{c}{$A_V$} & \multicolumn{1}{c}{$\log(Z/Z_\odot)$} \\
\hline
EGS-25297$^{1}$            & 9.94 & $-20.00^{+0.00}_{-0.00}$   & $-1.51^{+0.08}_{-0.08}$   & $117.00^{+21.00}_{-18.00}$  & $8.95^{+0.04}_{-0.05}$  & $<170$ & $0.92^{+0.42}_{-0.27}$ & $-1.00^{+0.08}_{-0.08}$ \\
UNCOVER-26185$^{2}$        & 10.05 & $-18.83^{+0.07}_{-0.07}$   & $-2.29^{+0.06}_{-0.06}$   & $1.70^{+0.40}_{-0.40}$     & $8.23^{+0.07}_{-0.09}$  & $293^{+42}_{-42}$ & $0.20^{+0.09}_{-0.09}$ & $-1.30^{+0.08}_{-0.09}$ \\
CEERS2-7929$^{3}$          & 10.10 & $-19.30^{+0.00}_{-0.00}$   & $-1.70^{+0.10}_{-0.10}$   & $5.90^{+1.60}_{-2.60}$     & $8.50^{+0.30}_{-0.40}$  & $520^{+127}_{-127}$ & $0.14^{+0.29}_{-0.14}$ & $-$ \\
MACS0647-JD2$^{4}$         & 10.17 & $-19.50^{+0.20}_{-0.20}$   & $-2.60^{+0.10}_{-0.10}$   & $2.40^{+1.00}_{-1.00}$     & $7.50^{+0.10}_{-0.10}$  & $70^{+24}_{-24}$ & $0.01^{+0.00}_{-0.00}$ & $-0.90^{+0.09}_{-0.09}$ \\
UNCOVER-37126$^{5}$        & 10.25 & $-20.10^{+0.05}_{-0.05}$   & $-2.88^{+0.10}_{-0.10}$   & $9.60^{+4.60}_{-4.60}$     & $7.77^{+0.06}_{-0.06}$  & $61^{+6}_{-6}$ & $0.03^{+0.03}_{-0.03}$ & $-$ \\
CAPERS-COS-109917$^{6}$    & 10.27 & $-19.57^{+0.11}_{-0.11}$   & $-2.84^{+0.28}_{-0.28}$   & $3.10^{+0.70}_{-0.80}$     & $8.00^{+0.40}_{-0.30}$  & $-$            &$0.02^{+0.03}_{-0.01}$ & $-0.90^{+0.20}_{-0.30}$ \\
GS-z10-0$^{7}$             & 10.38 & $-18.61^{+0.10}_{-0.10}$   & $-2.49^{+0.22}_{-0.22}$   & $1.10^{+0.19}_{-0.16}$     & $7.58^{+0.19}_{-0.20}$  & $<62$ & $0.05^{+0.03}_{-0.02}$ & $-1.91^{+0.25}_{-0.20}$ \\
CAPERS\_UDS\_z10$^{8}$     & 10.56 & $-20.53^{+0.09}_{-0.09}$   & $-2.27^{+0.12}_{-0.12}$   & $18.00^{+5.00}_{-5.00}$    & $8.30^{+0.20}_{-0.20}$  & $420^{+70}_{-70}$ & $0.31^{+0.10}_{-0.10}$ & $-$ \\
GN-z11$^{9}$               & 10.60 & $-21.50^{+0.02}_{-0.02}$   & $-2.36^{+0.10}_{-0.10}$   & $18.78^{+0.81}_{-0.69}$    & $8.73^{+0.06}_{-0.06}$  & $64^{+20}_{-20}$ & $0.17^{+0.03}_{-0.03}$ & $-0.92^{+0.06}_{-0.05}$ \\
JADES-GS-20176151$^{6}$    & 10.62 & $-19.50^{+0.14}_{-0.14}$   & $-2.11^{+0.17}_{-0.17}$   & $2.70^{+0.80}_{-0.50}$     & $8.10^{+0.10}_{-0.20}$  & $-$ & $0.02^{+0.06}_{-0.01}$ & $-1.90^{+0.20}_{-0.20}$ \\
EGS-22637$^{10}$           & 10.70 & $-20.23^{+0.12}_{-0.12}$   & $-2.42^{+0.09}_{-0.09}$   & $3.54^{+1.34}_{-0.98}$     & $8.62^{+0.14}_{-0.14}$  & $524^{+134}_{-134}$ & $0.11^{+0.13}_{-0.13}$ & $-1.69^{+0.40}_{-10.0}$ \\
MoM-z11-2$^{10}$           & 10.80 & $-18.82^{+0.36}_{-0.36}$   & $-1.60^{+0.22}_{-0.22}$   & $2.69^{+1.02}_{-0.74}$     & $8.58^{+0.18}_{-0.18}$  & $555^{+234}_{-234}$ & $0.13^{+0.10}_{-0.10}$ & $-0.68^{+0.25}_{-0.62}$ \\
CAPERS-EGS-43539$^{10}$    & 10.82 & $-18.82^{+0.36}_{-0.36}$   & $-2.78^{+0.09}_{-0.09}$   & $0.26^{+0.18}_{-0.11}$     & $7.51^{+0.28}_{-0.28}$  & $<61$ & $0.21^{+0.17}_{-0.17}$ & $-1.52^{+0.30}_{-10.0}$ \\
JADES-GS-20177294$^{6}$    & 10.89 & $-19.25^{+0.09}_{-0.09}$   & $-2.34^{+0.55}_{-0.55}$   & $0.30^{+0.60}_{-0.30}$     & $8.10^{+0.20}_{-0.20}$  & $-$ & $0.07^{+0.11}_{-0.05}$ & $-1.30^{+0.50}_{-0.50}$ \\
MoM-z11-1$^{10}$           & 10.92 & $-19.98^{+0.24}_{-0.24}$   & $-2.29^{+0.36}_{-0.36}$   & $4.78^{+2.29}_{-1.55}$     & $8.96^{+0.18}_{-0.18}$  & $636^{+235}_{-235}$ & $0.26^{+0.15}_{-0.15}$ & $-1.52^{+0.60}_{-10.0}$ \\
CAPERS\_UDS\_z11$^{8}$     & 11.01 & $-20.28^{+0.08}_{-0.08}$   & $-1.71^{+0.12}_{-0.12}$   & $43.00^{+10.00}_{-10.00}$  & $8.70^{+0.20}_{-0.20}$  & $560^{+61}_{-61}$ & $0.74^{+0.11}_{-0.11}$ & $-$ \\
CEERS2-588$^{11}$          & 11.04 & $-20.40^{+0.00}_{-0.00}$   & $-1.74^{+0.25}_{-0.25}$   & $8.20^{+0.00}_{-0.00}$     & $9.05^{+0.07}_{-0.09}$  & $453^{+120}_{-82}$ & $0.12^{+0.10}_{-0.12}$ & $-0.14^{+0.10}_{-0.13}$ \\
JADES-GS-20015720$^{6}$    & 11.27 & $-19.56^{+0.17}_{-0.17}$   & $-2.67^{+0.12}_{-0.12}$   & $2.90^{+0.70}_{-1.00}$     & $8.20^{+0.40}_{-0.30}$  & $-$ & $0.04^{+0.06}_{-0.03}$ & $-1.50^{+0.40}_{-0.30}$ \\
GS-z11-1$^{12}$            & 11.25 & $-19.40^{+0.08}_{-0.08}$   & $-2.80^{+0.10}_{-0.10}$   & $0.32^{+0.11}_{-0.11}$     & $7.80^{+0.20}_{-0.20}$  & $73^{+10}_{-10}$ & $0.05^{+0.03}_{-0.02}$ & $-1.55^{+0.12}_{-0.13}$ \\
MAISIE$^{13}$              & 11.44 & $-20.32^{+0.08}_{-0.06}$   & $-2.47^{+0.09}_{-0.09}$   & $2.10^{+4.80}_{-2.00}$     & $8.50^{+0.29}_{-0.44}$  & $340^{+14}_{-14}$ & $0.07^{+0.09}_{-0.05}$ & $-$ \\
EGS-z11-R0$^{14}$          & 11.45 & $-19.78^{+0.17}_{-0.17}$   & $-0.68^{+0.30}_{-0.30}$   & $3.86^{+0.76}_{-0.76}$     & $8.64^{+0.32}_{-0.32}$  & $<280$ & $0.90^{+0.28}_{-0.28}$ & $-$ \\
GS-z11-0$^{7}$             & 11.58 & $-19.34^{+0.05}_{-0.05}$   & $-2.18^{+0.09}_{-0.09}$   & $2.20^{+0.28}_{-0.22}$     & $8.67^{+0.08}_{-0.13}$  & $77^{+8}_{-8}$ & $0.18^{+0.06}_{-0.06}$ & $-1.87^{+0.28}_{-0.18}$ \\
GHZ2$^{15}$                & 12.34 & $-20.49^{+0.01}_{-0.01}$   & $-2.39^{+0.07}_{-0.07}$   & $5.20^{+1.10}_{-0.60}$     & $9.05^{+0.10}_{-0.25}$  & $105^{+9}_{-9}$ & $0.04^{+0.07}_{-0.03}$ & $-1.40^{+0.27}_{-0.24}$ \\
UNCOVER-z12$^{16}$         & 12.39 & $-19.20^{+0.50}_{-0.50}$   & $-2.34^{+0.30}_{-0.30}$   & $2.15^{+0.81}_{-0.46}$     & $8.35^{+0.14}_{-0.18}$  & $426^{+40}_{-42}$ & $0.19^{+0.17}_{-0.10}$ & $-1.34^{+0.60}_{-0.42}$ \\
GS-z12-0$^{7}$             & 12.63 & $-18.23^{+0.16}_{-0.16}$   & $-1.84^{+0.19}_{-0.19}$   & $1.80^{+0.54}_{-0.43}$     & $7.64^{+0.66}_{-0.39}$  & $144^{+15}_{-15}$ & $0.19^{+0.17}_{-0.10}$ & $-1.44^{+0.23}_{-0.22}$ \\
GS-z13-1-LA$^{17}$         & 13.01 & $-18.49^{+0.04}_{-0.04}$   & $-2.75^{+0.10}_{-0.10}$   & $0.16^{+0.48}_{-0.15}$     & $7.74^{+0.52}_{-0.18}$  & $14^{+14}_{-7}$ & $0.04^{+0.05}_{-0.03}$ & $-2.52^{+0.57}_{-0.48}$ \\
UNCOVER-z13$^{16}$         & 13.08 & $-19.40^{+1.80}_{-1.80}$   & $-2.59^{+0.20}_{-0.20}$   & $1.28^{+0.27}_{-0.18}$     & $8.13^{+0.11}_{-0.15}$  & $309^{+110}_{-74}$ & $0.04^{+0.08}_{-0.03}$ & $-1.57^{+0.35}_{-0.28}$ \\
GS-z13-0$^{7}$             & 13.20 & $-18.73^{+0.06}_{-0.06}$   & $-2.37^{+0.12}_{-0.12}$   & $1.36^{+0.31}_{-0.23}$     & $7.95^{+0.19}_{-0.29}$  & $<52$ & $0.04^{+0.08}_{-0.03}$ & $-1.69^{+0.28}_{-0.31}$ \\
PAN-z14-1$^{1}$            & 13.53 & $-20.60^{+0.20}_{-0.20}$   & $-2.26^{+0.08}_{-0.08}$   & $4.80^{+13.00}_{-4.80}$    & $8.23^{+1.14}_{-0.21}$  & $233^{+10}_{-10}$ & $0.06^{+0.18}_{-0.30}$ & $-1.40^{+1.21}_{-0.60}$ \\
GS-z14-1$^{18}$            & 13.90 & $-19.00^{+0.40}_{-0.40}$   & $-2.71^{+0.19}_{-0.19}$   & $2.00^{+0.70}_{-0.40}$     & $8.00^{+0.40}_{-0.30}$  & $<160$ & $0.20^{+0.11}_{-0.07}$ & $-1.10^{+0.60}_{-0.50}$ \\
GS-z14-0$^{18}$            & 14.18 & $-20.81^{+0.16}_{-0.16}$   & $-2.20^{+0.07}_{-0.07}$   & $19.00^{+6.00}_{-6.00}$    & $8.84^{+0.09}_{-0.10}$  & $260^{+2}_{-2}$ & $0.31^{+0.14}_{-0.17}$ & $-0.78^{+0.03}_{-0.03}$ \\
MoM-z14$^{19}$             & 14.44 & $-20.20^{+0.00}_{-0.00}$   & $-2.47^{+0.17}_{-0.17}$   & $13.00^{+3.70}_{-3.50}$    & $8.10^{+0.30}_{-0.20}$  & $74^{+15}_{-12}$ & $0.20^{+0.20}_{-0.10}$ & $-1.38^{+0.65}_{-0.56}$ \\
\hline
\end{tabular}
\vspace{0.3cm}

\parbox{\textwidth}{\small
{\bf Legend:} $^{1}$\citet[]{Donnan26}, $^{2}$\citet[]{Alvarez26}, $^{3}$\citet[]{Harikane24}, $^{4}$\citet[]{Hsiao23}, $^{5}$\citet[]{Marques26}, $^{6}$\citet[]{Tang26}, $^{7}$\citet[]{Curtis23}, $^{8}$\citet[]{Kokorev25}, $^{9}$\citet[]{Bunker23}, $^{10}$\citet[]{Roberts25}, $^{11}$\citet[]{Harikane26}, $^{12}$\citet[]{Scholtz25}, $^{13}$\citet[]{Finkelstein24}, $^{14}$\citet[]{Rodighiero26}, $^{15}$\citet[]{Castellano24}, $^{16}$\citet[]{Wang23}, $^{17}$\citet[]{Witstok25}, $^{18}$\citet[]{Carniani24a}, $^{19}$\citet[]{Naidu25}.
}
\end{table*}

\subsection{Star formation rate}\label{subsec:SFR}
In the AFM, the mean star formation rate (SFR) in a halo of total mass $M$ at redshift $z$ is  
\be
{\rm SFR} (M,z) = \epsilon_\star(v_c) f_b \frac{M}{t_{\rm ff}(z)},
\label{eq:SFR}
\ee
where $\epsilon_\star$ is the \textit{instantaneous} star formation efficiency depending on the halo circular velocity, $v_c(M,z)$; $f_b = \Omega_b/\Omega_m = 0.158$. The gas free-fall time in halos, $t_{\rm ff} = (4\pi G \rho)^{-1/2}$, can be conveniently written as $t_{\rm ff}= \zeta H(z)^{-1}$, where $H(z)^{-1}$ is the Hubble time at $z$ and $\zeta=0.06$ \citep[]{Ferrara24b}. 

Following \citet[]{Ferrara23} we write the efficiency as  
\begin{equation}
\epsilon_\star(v_c)  = \epsilon_0 \frac{v_c^2}{v_c^2 + f_w v_s^2},
\label{eq:eps}
\end{equation}
where $f_w = 0.12$ is the outflow coupling constant with the gas, and $v_s = \sqrt{\nu E_0}$ is a characteristic velocity associated with the SN energy ($E_0=10^{51}\ \rm erg$) released per unit stellar mass formed. We adopt $1-100\, \msun$ Salpeter IMF for which $\nu = (52.9\ M_\odot)^{-1}$, and $v_s = 975\ {\rm km\ s}^{-1}$. According to eq. \ref{eq:SFR}, the efficiency grows with $v_c^2$ (or $M_{\rm vir}^{2/3}$) at low masses and flattens to $\epsilon_0=0.09$ for circular velocities $v_c \simgt f_w^{1/2} v_s$.

The previous incarnation of AFM and its parameters successfully reproduces the UV LF up to $z\sim 14$, along with the cosmic star formation rate density \citep[]{Ferrara24a} and UV slope evolution \citep[]{Ferrara25c} and other observed super-early galaxy properties \citep[]{Ferrara24b, Ferrara25b}. In Fig. \ref{fig:AFM_UVLF} we show a comparison of the predicted UV LF at $z=11$ and $z=12$ with the available data. AFM matches very well the data at both redshifts. Interestingly, the shape of the predicted UV LF is perfectly fit by a standard double-power law whose parameters are reported in the two panels. 
The faint-end slope at these epoch is rather steep, $\alpha \simeq -2.1$, when computed in the range $-22 < M_{\rm UV} < -17$; we note that the slope is somewhat dependent on this choice.  
AFM also shows an excellent agreement to the cosmic star formation rate density data in the redshift range $5 < z <15$. However, we note that some tension is appearing for $z \simgt 14$ where the data seem to indicate a slower decline. Moreover, if the photometric candidates \citep[]{Perez25, Castellano25} at $z\simeq 17$ and $25$ are confirmed they would be extremely hard to reconcile not only with AFM but with essentially all $\Lambda$CDM-based galaxy evolution models. In that case one would be forced to introduce significant modifications, as e.g. PBH powering these sources \citep[]{Matteri25, Liu23}.   
Barring these uncertain data, we consider the agreement of the data with AFM as robust.

Using the observed values of $r_e$ and SFR for each galaxy in the sample in combination with eqs. \ref{eq:re}--\ref{eq:SFR}, we obtain\footnote{We account for errors on the free parameters of the model by sampling the experimental errors whose distribution is assumed to be gaussian.}  $\lambda$ and $v_c$, which can be used to obtain the halo mass $M$, the corresponding amplitude of the fluctuation in units of the r.m.s. of the cosmological density field, $\nu_\delta = \delta/ \sigma$, the halo number density, $n_M$, and the star formation efficiency $\epsilon(v_c)$. {Here $n_M\equiv dn/d\ln M$ is the differential halo mass function evaluated at the inferred halo mass and redshift
of each galaxy.}
These quantities are reported in Tab. \ref{tab:predicted_props}.

\subsection{Stellar mass}\label{subsec:Mstar}
The stellar mass is obtained by integrating eq. \ref{eq:SFR}, 
\begin{equation}\label{eq:Mstar0}
M_\star(t) = (1-R)\int_{t_*}^t \epsilon_\star(v_c) f_b\frac{M}{t_{\rm ff}(z)} \bigg|\frac{dt}{dz}\bigg| dz.
\end{equation}
where $R$ is the gas return fraction of stars for the adopted IMF. Here we will make the simplifying assumption that most of the stars are formed in the last free-fall time i.e. starting at $t_*= t-t_{\rm ff}$.
{We parameterize the duration of the main stellar mass-assembly episode as
$\Delta_{t_\star} t_{\rm ff}$, where $\Delta_{ t_\star}$ is a dimensionless
factor that is determined for each galaxy by requiring eq.~\ref{eq:Mstar0}
to reproduce its observed stellar mass. Thus, $\Delta_{t_\star}$ is not
fixed to unity; for the galaxies analysed below it spans a considerably
wider range. The quantity $\Delta_{t_\star} t_{\rm ff}$ should therefore
be interpreted as an effective mass-assembly timescale rather than as
a precise stellar age obtained from a fitted star-formation history.}

{For $\Delta t_\star\simeq1$ at $z\simeq10$,
the corresponding mass-assembly timescale is
$t_{\rm a}\simeq40$ Myr, comparable in magnitude
to stellar ages inferred from some non-parametric
SFH analyses
\citep{Bouwens22b,Whitler23,Donnan24,Lopez26},
although the two quantities are not generally equivalent.}

Given such a short time lapse, we approximate the stellar mass as
\be
M_*(M,z) \simeq  (1-R) \epsilon_\star(v_c) f_b \Delta_{t_*} M;
\label{eq:Mstar}
\ee
it also follows that $R$ is much smaller than in the commonly adopted case of Instantaneous Recycling Approximation (IRA), according to which stars with mass $m < m_0$ ($m \ge m_0$) live forever (die immediately). Thus, for simplicity, we set $(1-R) \approx 1$ for numerical estimates. As all the other quantities are known, we use eq. \ref{eq:Mstar} and the observed value of $M_*$ for each galaxy to derive $\Delta_{t_*}$.
\subsection{Specific star formation rate}\label{subsec:sSFR}
Armed with the above expressions for the SFR (eq. \ref{eq:SFR}) and $M_*$ (eq. \ref{eq:Mstar}), it is straightforward to write the specific star formation rate, sSFR = SFR$/M_*$, as
\be
{\rm sSFR} = 0.64\,\frac{(1+z)^{3/2}}{(1-R)\Delta_{t_*}}\, \rm Gyr^{-1}.
\label{eq:sSFR}
\ee
We note that the sSFR is a function of redshift only and grows monotonically. {For $\Delta_{t_*}\simeq1$, this redshift scaling reproduces well the
available population-level data and existing models (see
\citealt{Ferrara24a,Santini25} for detailed comparisons).}

\subsection{Super-Eddington condition} \label{subsec:SupEdd}
The basic idea behind AFM is that galaxies can go through evolutionary phases in which their bolometric luminosity, $L_{\rm bol}$,  becomes super-Eddington,
\begin{equation}
L_{\rm bol} > L_E^{\rm eff} = A^{-1} L_E, 
\label{eq:superEdd}
\end{equation}
where $L_{E}= (4\pi G m_p c /\sigma_T) M_* = 1.26\times 10^{38} (M_*/M_\odot)\, \rm erg\, s^{-1}$ is the classical Eddington luminosity. This is reduced by a \quotes{boost} factor $A \approx \sigma_d/\sigma_T$, i.e. the ratio of the dust-to-Thomson cross-section in a dusty medium. \citet{Nakazato25} find that 
\be
A \simeq 1800 \left(\frac{Z}{Z_\odot}\right) \left[1+ \frac{N_{\rm H}}{10^{23.5}\rm cm^{-2}}\right]^{-1},
\label{eq:A}
\ee
where $Z$ is the gas metallicity (solar value $Z_\odot=0.0142$ \citealt{Asplund09}) and $N_{\rm H}$ the hydrogen column density in the galaxy. For simplicity, we will neglect the weak $N_{\rm H}$ dependence. 

As $L_{\rm bol}$ is proportional to the SFR and $L_E\propto M_*$, \citet{Ferrara24a} showed that the super-Eddington condition eq. \ref{eq:superEdd}  translates into one on the sSFR: 
\begin{equation}
{\rm sSFR} > {\rm sSFR}^\star \simeq 1.4 \left(\frac{Z_\odot}{Z}\right) \left(\frac{2}{f_{\rm bol}}\right) {\rm Gyr^{-1}},
\label{eq:ssfr_thresh}
\end{equation}
where $f_{\rm bol} \simeq 2$ is the bolometric to UV correction \citep[]{Fiore23}, and we have used eq. \ref{eq:A} to eliminate $A$. Eq. \ref{eq:ssfr_thresh} represents the necessary condition for a galaxy to develop a radiation-driven outflow displacing its dust and gas to larger radii.

{We stress that the super-Eddington classification used below is obtained
by comparing the observed ${\rm sSFR}={\rm SFR}/M_\star$ with
${\rm sSFR}^\star$ from eq.~\ref{eq:ssfr_thresh}. It therefore does
not depend on the inferred value of $\Delta t_\star$.
Eq.~\ref{eq:sSFR} follows algebraically from eq. \ref{eq:SFR} and \ref{eq:Mstar}, with $\Delta t_\star$ itself fixed by the observed $M_*$ and is not used as an additional constraint.}

The gas metallicity, i.e. the metal-to-gas mass ratio $Z=M_Z/M_g$, can be estimated from the IMF-averaged SN metal yield $y_Z = 2.41\ M_\odot$. 
Following \citet[]{Ferrara25b} this can be written in closed form as  
\begin{equation}
Z \simeq  y_Z\nu  \frac{(1-R)}{a(z,\epsilon_\star)}. 
\label{eq:metallicity}  
\end{equation}
The previous expression also accounts for the cosmological accretion via the term
\begin{equation}
a(z, \epsilon_\star) = 6.23[-a_0 -a_1 (1+z)]\left(\frac{0.01}{\epsilon_\star}\right); 
\label{eq:a}  
\end{equation}
where $(a_0, a_1)$ are parameters that depend on cosmology, and the linear matter power spectrum. These are provided\footnote{For reference, when averaged over $8 < \log (M_0/\msun) < 14$, $(a_0, a_1)=(0.25, -0.75)$.} in Appendix C of \citet{Correa15}.

Combining eqs. \ref{eq:ssfr_thresh} and \ref{eq:metallicity} leads to conclude that \textit{on average} galaxies (using fiducial values of the parameters) become super-Eddington for the first time when the star formation efficiency satisfies the inequality 
\be
\epsilon_\star(v_c) \simgt {0.04} \Delta_{t_*} (1+z)^{-1/2}. 
\label{eq:SEdd_condition}
\ee

%
%
\begin{sidewaystable*}
\centering
\scriptsize
\setlength{\tabcolsep}{2pt}
\noindent\begin{minipage}[c][\textheight][c]{\textwidth}
\centering
\caption{Predicted properties of super-early galaxies. Galaxies are coded as G1 to G17 (see legend below).}
\label{tab:predicted_props}

\makebox[\linewidth][c]{%
\begin{tabular}{l|*{17}{r}}
\hline\hline
Property          & G1 & G2 & G3 & G4 & G5 & G6 & G7 & G8 & G9 & G10 & G11 & G12 & G13 & G14 & G15 & G16 & G17 \\
\hline
$z$               & 9.94 & 10.05 & 10.10 & 10.56 & 10.70 & 10.80 & 10.92 & 11.01 & 11.04 & 11.44 & 11.45 & 12.39 & 12.63 & 13.08 & 13.53 & 13.90 & 14.18 \\
$r_e$ [pc]        & 170  & 293   & 520   & 420   & 524   & 555   & 636   & 560   & 453   & 340   & 280   & 426   & 144   & 309   & 233   & 160   & 260   \\
$\lambda$         & 0.0057 & 0.0259 & 0.0354 & 0.0237 & 0.0431 & 0.0491 & 0.0505 & 0.0274 & 0.0324 & 0.0343 & 0.0248 & 0.0478 & 0.0172 & 0.0417 & 0.0249 & 0.0214 & 0.0218 \\
$v_c\ [{\rm km\ s^{-1}}]$      & 344.14 & 131.72 & 171.97 & 220.35 & 153.98 & 145.19 & 164.28 & 269.63 & 184.83 & 137.74 & 156.87 & 138.43 & 133.32 & 124.08 & 164.43 & 136.32 & 223.07 \\
$\epsilon_*$   & 0.0458 & 0.0119 & 0.0185 & 0.0269 & 0.0155 & 0.0140 & 0.0172 & 0.0350 & 0.0207 & 0.0128 & 0.0160 & 0.0129 & 0.0121 & 0.0107 & 0.0172 & 0.0126 & 0.0273 \\
$\Delta_{t_*}$          & 0.18  & 2.35   & 1.27   & 0.28   & 3.02   & 3.67   & 5.03   & 0.31   & 3.66   & 4.23   & 3.18   & 3.27   & 0.78   & 3.57   & 1.26   & 1.84   & 1.38   \\
$r_{\rm out}/r_e$              & 4.77  & 7.40   & 3.24   & 1.68   & 8.63   & 9.25   & 9.64   & 1.24   & 14.55  & 24.47  & 4.93   & 6.70   & 3.53   & 22.15  & 8.14   & 7.30   & 5.63   \\
$\nu_\delta = \delta/\sigma$    & 6.12 & 4.32 & 4.75 & 5.38 & 4.78 & 4.71 & 4.96 & 5.99 & 5.22 & 4.84 & 5.06 & 5.15 & 5.16 & 5.18 & 5.85 & 5.60 & 6.75 \\
$\log (M/M_\odot)$ & 11.843 & 10.585 & 10.930 & 11.226 & 10.751 & 10.669 & 10.824 & 11.464 & 10.971 & 10.566 & 10.735 & 10.525 & 10.464 & 10.350 & 10.696 & 10.435 & 11.065 \\
$\log (M_*/M_\odot)$        & 8.950 & 8.230  & 8.500  & 8.300  & 8.620  & 8.580  & 8.960  & 8.700  & 9.050  & 8.500  & 8.640  & 8.350  & 7.640  & 8.130  & 8.230  & 8.000  & 8.840  \\
SFR  $[M_\odot\ {\rm yr^{-1}}]$             & 117.00 & 1.70  & 5.90   & 18.00  & 3.54   & 2.69   & 4.78   & 43.00  & 8.20   & 2.10   & 3.86   & 2.15   & 1.80   & 1.28   & 4.80   & 2.00   & 19.00  \\
sSFR $[{\rm Gyr^{-1}}]$               & 131.28 & 10.01 & 18.66  & 90.21  & 8.49   & 7.08   & 5.24   & 85.80  & 7.31   & 6.64   & 8.84   & 9.60   & 41.24  & 9.49   & 28.26  & 20.00  & 27.46  \\
sSFR$^*$ $[{\rm Gyr^{-1}}]$          & 6.28  & 24.48 & 15.77  & 11.34  & 19.92  & 22.17  & 18.26  & 9.05   & 15.32  & 25.60  & 20.59  & 27.37  & 29.74  & 34.86  & 22.34  & 31.35  & 14.74  \\
$A_{\rm UV}$      & 1.832 & 0.503 & 0.358  & 0.753  & 0.284  & 0.333  & 0.642  & 1.562  & 0.309  & 0.182  & 1.803  & 0.479  & 0.479  & 0.105  & 0.157  & 0.503  & 0.753  \\
$A_V$             & 0.920 & 0.200 & 0.140  & 0.310  & 0.110  & 0.130  & 0.260  & 0.740  & 0.120  & 0.070  & 0.900  & 0.190  & 0.190  & 0.040  & 0.060  & 0.200  & 0.310  \\
$\beta$           & -1.243 & -2.258 & -2.343 & -2.103 & -2.385 & -2.357 & -2.173 & -1.497 & -2.371 & -2.441 & -1.271 & -2.272 & -2.272 & -2.484 & -2.455 & -2.258 & -2.103 \\
$M_{\rm UV}^{\rm int}$      & -23.690 & -19.096 & -20.447 & -21.658 & -19.893 & -19.594 & -20.219 & -22.604 & -20.805 & -19.326 & -19.986 & -19.351 & -19.158 & -18.788 & -20.223 & -19.273 & -21.717 \\
$M_{\rm UV}$ & -21.859 & -18.594 & -20.089 & -20.905 & -19.609 & -19.261 & -19.577 & -21.042 & -20.496 & -19.143 & -18.183 & -18.872 & -18.679 & -18.683 & -20.066 & -18.770 & -20.963 \\
$Z\ [Z_{\odot}]$     & 0.2230 & 0.0572 & 0.0888 & 0.1235 & 0.0703 & 0.0632 & 0.0767 & 0.1548 & 0.0914 & 0.0547 & 0.0680 & 0.0511 & 0.0471 & 0.0402 & 0.0627 & 0.0447 & 0.0950 \\
$\log[(dn/d\ln M)/{\rm Mpc}^{-3}]$& -7.61 & -3.67 & -4.57 & -5.77 & -4.43 & -4.27 & -4.76 & -7.01 & -5.28 & -4.34 & -4.82 & -4.77 & -4.73 & -4.65 & -6.05 & -5.40 & -8.08 \\
\hline
\end{tabular}%
}
\vspace{0.3cm}

\parbox{\textwidth}{\centering
\textbf{Legend (galaxy codes to original names):}\\
\begin{tabular}{lllll}
G1 = EGS-25297 & G2 = UNCOVER-26185 & G3 = CEERS2-7929 & G4 = CAPERS\_UDS\_z10 & G5 = EGS-22637 \\
G6 = MoM-z11-2 & G7 = MoM-z11-1 & G8 = CAPERS\_UDS\_z11 & G9 = CEERS2-588 & G10 = MAISIE \\
G11 = EGS-z11-R0 & G12 = UNCOVER-z12 & G13 = GS-z12-0 & G14 = UNCOVER-z13 & G15 = PAN-z14-1 \\
G16 = GS-z14-1 & G17 = GS-z14-0
\end{tabular}
}
\end{minipage}
\vspace{6cm}

\end{sidewaystable*}

\subsection{Dust obscuration} \label{sec:Obscured}
Having established when the galaxy enters the super-Eddington phase, we now aim at determining if and when the galaxy will become obscured during its evolution. As star formation proceeds, supernovae (SN) will inject freshly synthesized dust grains which attenuate the stellar light. Finally $\xi_d = M_d/M_* = y_d \nu/(1-R)$ is the production-only\footnote{As we are interested in redshift $z>10$ the timescale for the destruction processes tend to be larger than the Hubble time and can be neglected to a first approximation.} dust-to-stellar mass ratio. We follow \citet[]{Ferrara25b} and take $y_d=0.1 M_\odot$ for the SN dust yield. The dust optical depth at 1500\AA\ can be written as 
\be
\tau_{\rm UV} = \frac{\kappa_{\rm UV} \xi_d}{2\pi q r_e^2}M_*,
\label{eq:tauUV}
\ee
where $r_e$ and $M_*$ are given by eq. \ref{eq:re} and eq. \ref{eq:Mstar}, respectively; $q=(1,2)$ for (disk, spherical) geometry\footnote{To conservatively minimize the optical depth we assume $q=2$, also consistent with the quasi-spherical stellar shape of super-early simulated galaxies \citep[]{Kohandel25}.}, $\kappa_{\rm UV}=1.26\times 10^5\ \rm cm^2 g^{-1}$ and $\tau_{\rm UV} = 2.655 \tau_V$ are the dust mass absorption coefficient for a Milky Way extinction curve \citep{Weingartner01}. The factor 2.655 accounts for the differential attenuation between 1500\AA\ and the V-band for the adopted extinction curve. 

When the super-Eddington condition $\rm sSFR > sSFR^*$ condition is satisfied, a radiation-driven outflow develops. The dust and the gas, initially contained within $r_e$ are then displaced to a larger radius, $r_{\rm out} > r_e$. In this case, we replace $r_e$ with $r_{\rm out}$ in the previous formula to compute $\tau_{\rm UV}$.    
Finally, the V-band attenuation is computed from   
\begin{equation}
A_V = -2.5 \log \frac{1-e^{-\tau_V}}{\tau_V},
\label{eq:AV}
\end{equation}
as appropriate for a disk configuration in which stars and dust are intermixed \citep[]{Ferrara22a}. The observed (i.e. including dust attenuation) galaxy AB magnitude at 1500 \AA\ is obtained from the following expression:
\begin{equation}
M_{\rm UV} = 5.89 - 2.5 \log ({\cal K}_{1500} {\rm SFR}) + A_{\rm UV},
\label{eq:MUV}
\end{equation}
where the SFR-luminosity conversion factor ${\cal K}_{1500} = 0.587 \times 10^{10} {L_\odot}/(M_\odot {\rm yr}^{-1})$, following \citet[]{Ferrara24a}; $A_{\rm UV}$ is computed from eq. \ref{eq:AV} using the above relation between UV and visual optical depth. 

From the attenuation, $A_V$, we can also predict the value of the spectral UV slope $\beta$. Such conversion is uncertain as it depends on the properties of the dust, which are poorly understood at these very early epochs. 

{We derive the relation between attenuation and UV slope empirically
from the 32 galaxies in Table~\ref{tab:observed_props}, for which both
$A_V$ and $\beta$ estimates are available. An unweighted linear
least-squares fit gives
\begin{equation}
\beta = (-2.54\pm0.08) + (1.45\pm0.26) A_V ,
\label{eq:beta}
\end{equation}
with an rms scatter $\sigma_\beta=0.32$. We use this empirical relation
to map the observed $A_V$ into the corresponding UV slope; as stressed
above, $\beta$ obtained in this way is therefore an input-dependent
observable rather than an independent AFM prediction.}

{In summary, we use four observed quantities $r_e, {\rm SFR}, M_*, A_V$ as constraints to the AFM, which returns the values of additional 12 properties for each galaxy. Among these, two are input-dependent observables (the attenuated UV magnitude, $M_{\rm UV}$, and the UV slope, $\beta$); the metallicity, $Z$ is a prediction of the model. The other nine inferred quantities are measured only in a limited number of objects and/or cannot be experimentally determined. These are $\lambda, v_c, \epsilon_*, \Delta_{t_*}, r_{\rm out}, M, \nu_\delta, {\rm sSFR}^*, M_{\rm UV}^{\rm int}$ (the unattenuated UV magnitude), and $n_M$ which are then used to test the outflow-based AFM scenario and its implications for the evolution of super-early galaxies.}  

{A quantitative comparison with the observed values is presented in
Appendix~\ref{app:residuals}, where we report the residuals in
$M_{\rm UV}$, $\beta$, and metallicity for the 17 galaxies included
in the detailed analysis. This comparison should be regarded as a
consistency test for the two input-dependent observables, $M_{\rm UV}$
and $\beta$, whereas metallicity provides a more direct test of the
model.}

{We caution that the SFR and stellar masses compiled
in Tab.~1 originate from heterogeneous analyses adopting
different assumptions about stellar populations,
star-formation histories, and dust attenuation.
Moreover, these quantities are often inferred from
overlapping photometric and spectroscopic information,
particularly the rest-frame UV continuum, and their
uncertainties may therefore be correlated. Such
systematics propagate directly into the inferred
${\rm sSFR}={\rm SFR}/M_\star$ and hence into the
super-Eddington classification.}

\section{Results} \label{sec:results}
{We apply the AFM model described above to the sample of 32
spectroscopically confirmed $z>10$ galaxies whose observed properties
are listed in Table~\ref{tab:observed_props}. For all of them the
redshift, UV magnitude, UV spectral slope, star formation rate, and
stellar mass are available, whereas effective-radius measurements are
not available for all sources.
For the detailed AFM analysis we select 17 galaxies on the basis of
their measured or adopted effective radii, using $r_e\simeq150$ pc as
the nominal boundary between the extended and compact populations.
This subset corresponds exactly to the 17 systems listed in
Table~\ref{tab:predicted_props}. We include GS-z12-0
($r_e=144\pm15$ pc) as a borderline case because its uncertainty
overlaps the adopted size threshold. For sources with an upper limit
above the threshold (EGS-25297, EGS-z11-R0, and GS-z14-1), we adopt
the quoted upper limit as the characteristic radius for the purpose
of the AFM calculation; as discussed below, the resulting solutions
should therefore be regarded as conditional on this assumption.
Of the remaining 15 galaxies, 11 have measured or constrained
$r_e\lesssim150$ pc and are discussed separately as the compact
population in Sec.~\ref{sec:compact}, while four sources
(CAPERS-COS-109917, JADES-GS-20176151, JADES-GS-20177294, and
JADES-GS-20015720) lack an effective-radius constraint and are
therefore not assigned to either size class. Thus, all quantitative
AFM-derived properties and the evolutionary analysis presented below
refer specifically to the 17-object subset in
Table~\ref{tab:predicted_props}.}

%
%
\begin{figure*}
\centering\includegraphics[width = 0.98 \textwidth]{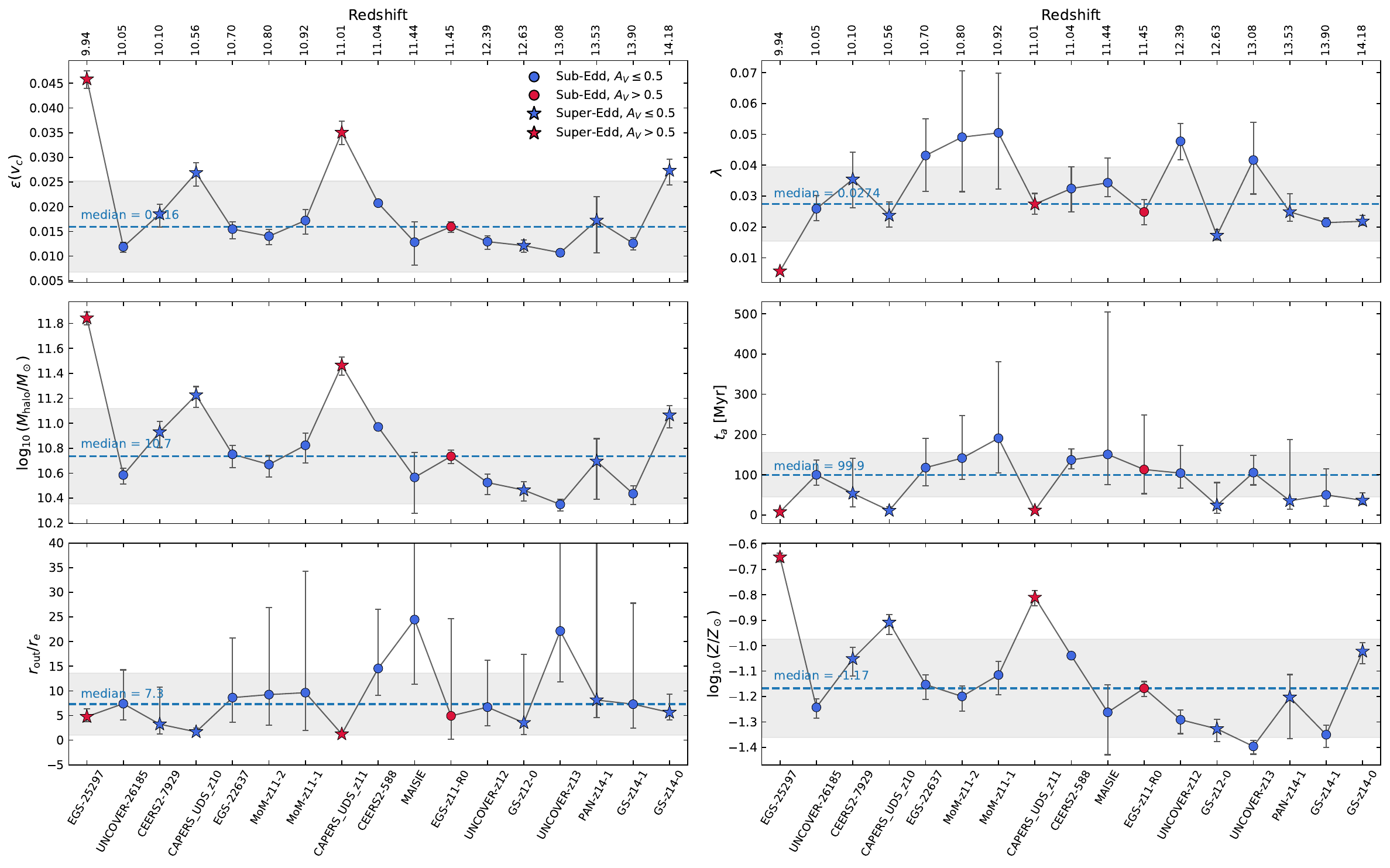}
\caption{Properties of super-early galaxies predicted by AFM. For the 17 galaxies selected for the detailed AFM analysis
(ordered according to redshift, shown on the top axis) in the sample we show the instantaneous star formation efficiency (top left panel), spin parameter (top right), halo mass (middle left), {effective mass-assembly timescale}  (middle right), outflow outer radius in units of the effective radius (bottom left), metallicity (bottom right). Data points are labelled with different markers/colors according to their Eddington parameter and visual attenuation (see legend). The horizontal line (bue dashed) shows the median of the sample, with the grey area spanning the 16th-84th percentile range.} 
\label{fig:properties}
\end{figure*}

\subsection{Physical properties} \label{sec:props}
{The predicted properties of the 17 galaxies selected for the detailed
AFM analysis are summarized in Fig.~\ref{fig:properties} and
Table~\ref{tab:predicted_props}}, where they are ordered in increasing redshift sequence\footnote{{For EGS-25297, EGS-z11-R0, and GS-z14-1, only upper limits on the
effective radius are available ($r_e<170$, $<280$, and $<160$ pc,
respectively). 
For these sources we adopt the quoted upper limit when deriving
the AFM quantities in Table~\ref{tab:predicted_props}. Their inferred
properties should therefore be regarded as conditional limiting
solutions rather than estimates based on a measured galaxy size.}}. 
We show six quantities: the instantaneous star formation efficiency, spin parameter, halo mass, {mass-assembly timescale}, outflow outer radius in units of the effective radius, and metallicity (bottom right). These quantities are not yet directly observable (metallicity is a partial exception).

In each panel we also highlight whether the galaxy is super-Eddington (and therefore potentially able to launch an outflow), and whether it is obscured. We adopt $A_V = 0.5$, corresponding to $\tau_V \simeq 1$, as the threshold between the transparent and obscured regime. 
Additional predictions not shown here can be found in Tab.~\ref{tab:predicted_props}. In the following we discuss the results and their implications separately for each quantity. 

\subsubsection{Star formation efficiency} \label{subsec:sfe}
The first notable finding is that the star formation efficiency is moderate, {ranging from $0.01 \lesssim \epsilon_* \lesssim 0.05$} with a median value of $0.016$. These results are in excellent agreement with those shown in Fig.~3 of \citet[]{Harikane26}, who find that $\epsilon_* < 0.03$ for all super-early galaxies, with the only exception of CEERS2-588, which appears to have a somewhat higher efficiency ($\epsilon_*\simeq 0.1$) compared to our prediction for the same system ($\epsilon_*=0.02$). 

We note, though, that \citet[]{Harikane26} obtain $\epsilon_*$ assuming a fixed halo mass $\log (M/M_\odot) = 10.7$ (the most massive halo in the CEERS survey volume, and in perfect agreement with the median $M$ value found here; see Sec.~\ref{subsec:halomass}), while for CEERS2-588 we derive a slightly larger value $M = 10^{10.97}\ \Msun$. Adopting our value for the halo mass, the observed efficiency would drop to $\epsilon_* = 0.054$, which is in better -- though not yet perfect -- agreement with our result. 

Finally, we note that the values inferred by \citet[]{Harikane26} refer to the \textit{integrated} star formation efficiency (i.e., the gas-to-star conversion efficiency) rather than the \textit{instantaneous} efficiency we use here. This difference may also contribute to the discrepancy.

No significant variations with redshift are seen. However, we note that the two red, super-Eddington galaxies (EGS-25297 and CAPERS\_UDS\_z11) lie $2$--$3\,\sigma$ away from the median value. In Sec.~\ref{sec:evol} we will interpret this evidence in the framework of an evolutionary scenario. 

\subsubsection{Spin parameter}\label{subsec:spin}
The second notable finding concerns the halo spin parameter \(\lambda\). Our inferred values span a range \(0.006 < \lambda < 0.05\) with a median of \(\lambda = 0.028\). These values are consistent with typical spin parameters predicted for dark matter halos in cosmological simulations, which peak around $\lambda \simeq 0.03-0.04$ \citep{Bullock01, Maccio07, Dutton14, Angel16}, and show no significant trend with redshift, as also found here.

Again, an exception to this trend is provided by EGS-25297, which shows a lower spin parameter, \(\lambda \approx 0.006\), lying more than \(2\sigma\) below the median value. Low-spin haloes are characterised by deeper potential wells and more concentrated mass distributions, which may hinder the launching of large-scale outflows and allow dust to remain in the galaxy centre, thereby producing the redder colours observed in these systems. This interpretation is consistent with the evolutionary scenario outlined in Sec.~\ref{sec:evol}.

\subsubsection{Halo mass}\label{subsec:halomass}
{AFM predicts that Blue Monsters are hosted by relatively massive halos,
with masses spanning approximately $10.3\lesssim\log(M/M_\odot)
\lesssim11.2$, and a median value $\log(M/M_\odot)=10.7$.
The full 17-object sample extends to still larger halo masses,
reaching $\log(M/M_\odot)=11.84$ for the Red Monster EGS-25297.} 
{Using the \citet[]{Eisentein98} smooth, baryon-inclusive linear transfer
function, normalized to $\sigma_8=0.826$, and a real-space top-hat
filter, we infer peak heights $\nu_\delta=\delta_c/[D(z)\sigma(M,0)]$
in the range $4.32$--$6.75$ (median $5.16$). The halo abundances
in Tab.~\ref{tab:predicted_props} are evaluated consistently using
the Sheth--Tormen mass function and are quoted as the differential
quantity $dn/d\ln M$.}
We also see a hint of decreasing halo mass towards high redshifts. 

Red monsters in a super-Eddington phase like EGS-25297 and CAPERS\_UDS\_z11 live instead on even more massive halos with $\log M \simgt 11.5$. Interestingly, the newly discovered $z=11.45$ Red Monster EGS-z11-R0 is instead in a sub-Eddington phase and its halo mass is the median one. Later on, we will see that this galaxy is on a similar evolutionary path as the other two Red Monsters, but it is seen at an earlier evolutionary phase. 

{As an additional abundance check, we integrate the same
Sheth--Tormen mass function over halo masses above each inferred
mass. The rarest systems are EGS-25297 and GS-z14-0, for which
$\log[n(>M)/{\rm Mpc}^{-3}]=-8.28$ and $-8.78$, respectively.
For GS-z14-0, the $58\,{\rm arcmin}^2$ JADES GOODS-S discovery area
\citep{Carniani24a}, together with the representative interval
$13.5<z<15$, gives a comoving volume
$V\simeq1.1\times10^5\,{\rm Mpc}^3$ and hence an expected number
$nV\simeq1.8\times10^{-4}$ of halos above the inferred mass.}

{The inferred rarity of these halos should be interpreted in the context of the assumptions underlying the AFM mass inference. In particular, halo masses are obtained primarily by inverting the adopted SFR–halo mass relation, while the observed effective radii constrain the corresponding spin parameters through eq. (1). Systematic uncertainties in the star-formation efficiency, UV-to-SFR conversion, and possible stochastic fluctuations in star formation could therefore lower the inferred halo masses and substantially alleviate the abundance tension.}

%
%
\begin{figure*}
\centering\includegraphics[width = 0.98 \textwidth]{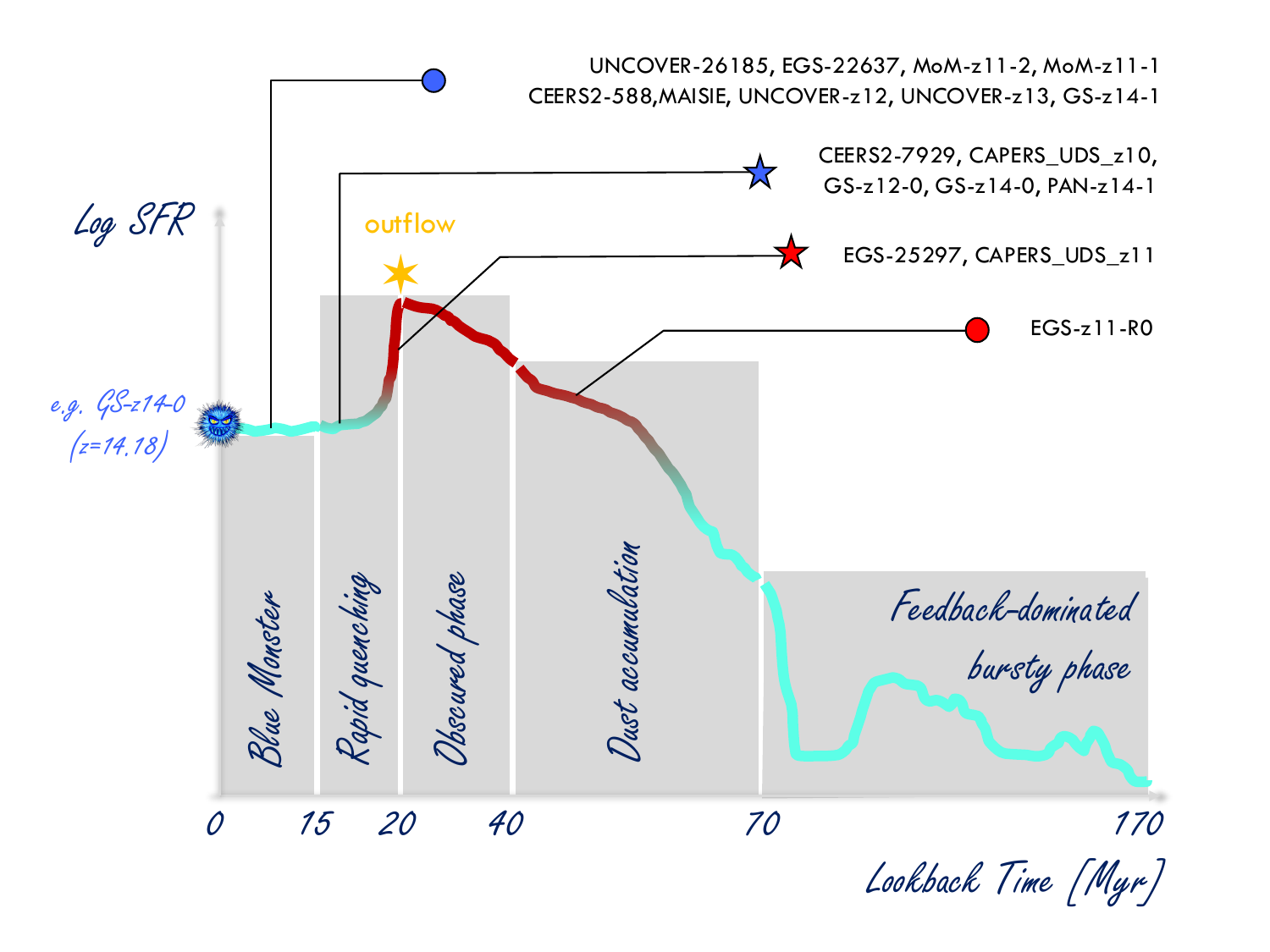}
\caption{Sketch of the possible evolutionary path leading to the formation of a Blue Monster in the Attenuation-Free Model (AFM), based on the detailed study of GS-z14-0 at $z=14.18$ presented in \citet[]{Ferrara24b}. The values of the lookback time and of the SFR are only indicative, as they depend on the properties of individual galaxies, as discussed in the text.} 
\label{fig:evo_scheme}
\end{figure*}
%

%
%
\begin{figure*}
\centering\includegraphics[width = 0.49 \textwidth]{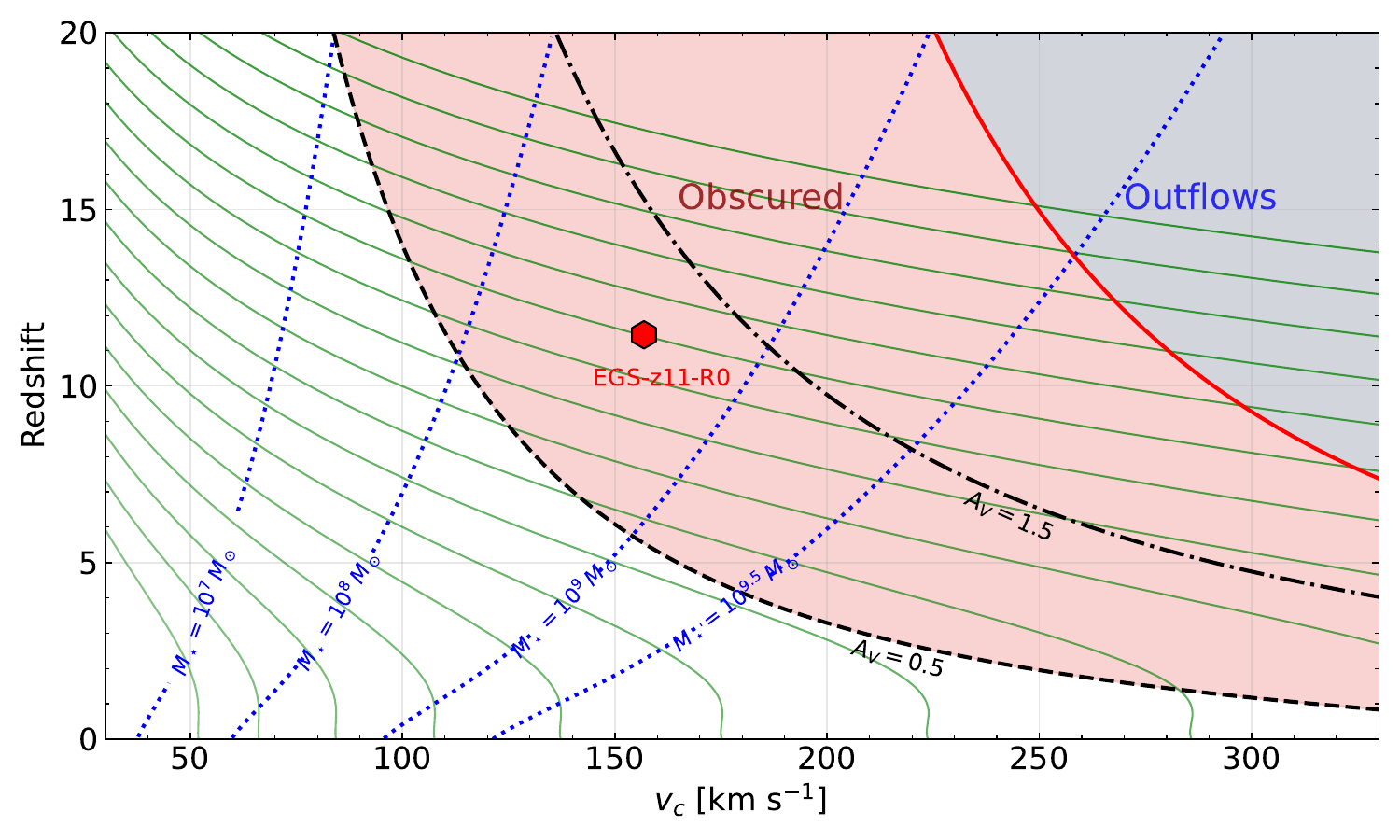}
\centering\includegraphics[width = 0.49 \textwidth]{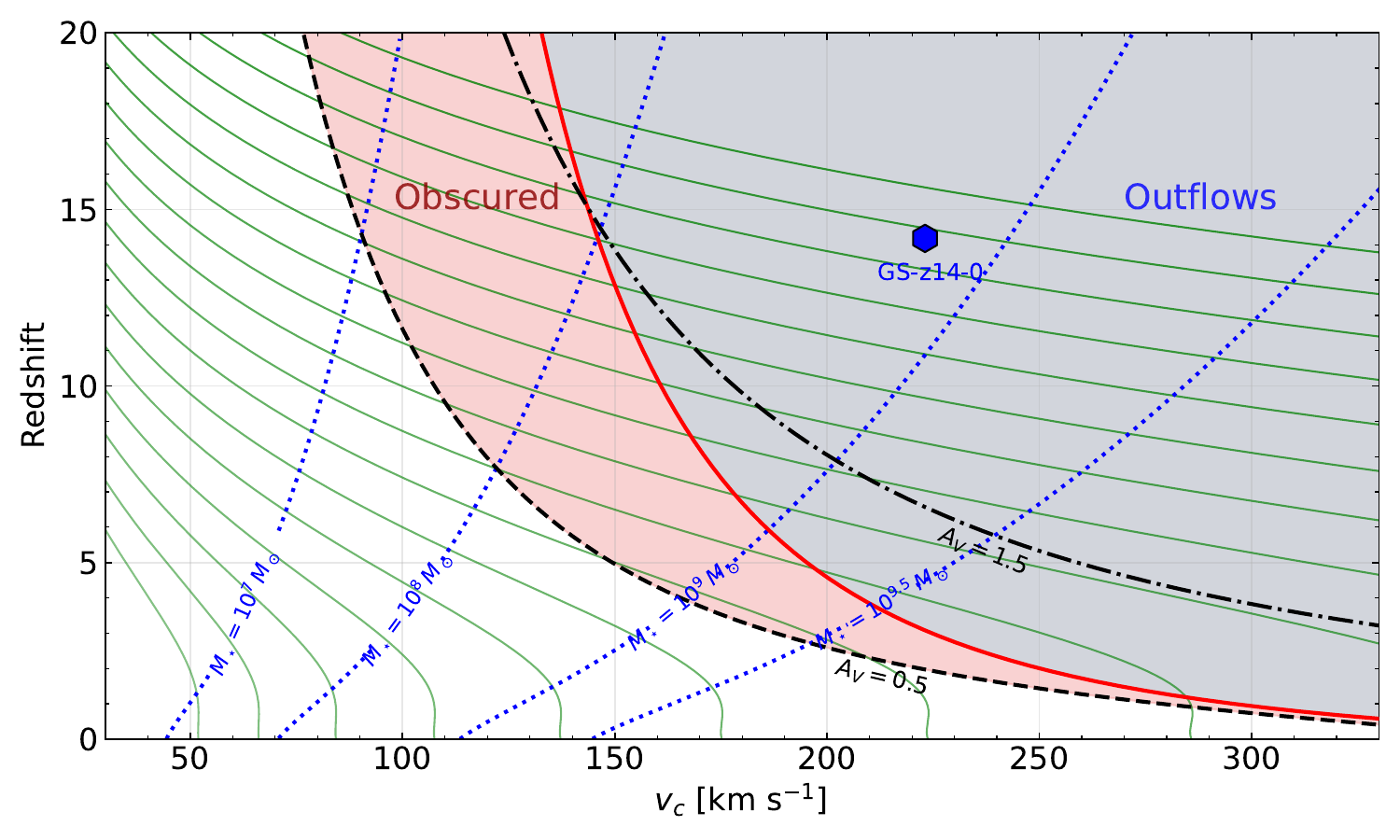}
\caption{Evolutionary path of super-early galaxies shown in the redshift - circular velocity (proxy for halo mass) plane. The green lines  (bottom to top) are the tracks of galaxies whose star formation activity starts at $z_* = 9,10, .., 25$, respectively. Also shown are stellar mass isocountours (blue dotted). As the circular velocity, halo and stellar mass increase with time the galaxy accumulates dust and becomes redder, crossing the $A_V = 0.5$ (dashed) line representing our definition of \quotes{obscured} systems; for reference we also plot the $A_V=1.5$ (dot-dashed) curve. When the track crosses the sSFR$^*$ threshold (red line) the galaxy becomes super-Eddington, and launches an outflow, eventually displacing the dust and turning it into a Blue Monster. \textit{Left panel}: Case for the Red Monster EGS-z1--R0 ($z=11.45$). \textit{Right}: Same for the Blue Monster GS-z14-0 ($z=14.18$).} 
\label{fig:z_vs_vc}
\end{figure*}

\subsubsection{Effective mass-assembly timescales}
\label{subsec:age}

{The effective mass-assembly timescale,
$t_{\rm a}\equiv\Delta t_\star t_{\rm ff}
=M_\star/{\rm SFR}$ (for $R=0$), is shown in the middle
right panel of Fig.~\ref{fig:properties}. It spans
$\simeq10-190$ Myr across the sample, with a median
of $99.9$ Myr. As this quantity is directly determined
by the observed SFR and stellar mass, it is not an
independent AFM prediction, nor does it uniquely
constrain the onset of star formation or the
mass-weighted stellar age. In particular, the short
$t_{\rm a}\simeq 10$ Myr values of the two
super-Eddington Red Monsters simply reflect their
high observed sSFRs, rather than demonstrating that
their stellar populations are exceptionally young.}

\subsubsection{Outflow radius}\label{subsec:outflow}
One of the key features of AFM is that it predicts that super-early galaxies, due to their compactness and astounding UV luminosity, can drive powerful radiative outflows displacing the dust and gas up to a radius $r_{\rm out}$ significantly larger than the effective radius $r_e$. The ratio of these quantities for the galaxies in the sample is shown in the bottom left panel of Fig.~\ref{fig:properties}.

On average, Blue Monsters have been able to push the dust to $7.3\,r_e$, thus decreasing the dust optical depth (see Eq.~\ref{eq:tauUV}). Some of the outflows are still ongoing (those in galaxies that are still in the super-Eddington phase). These systems tend to have smaller outflow radii as the outflow has not fully developed yet. Among the systems in this phase are GS-z12-0 and GS-z14-0 (see blue stars in the panel). 

Other Blue Monsters are instead sub-Eddington, and the outflow has run out of power, reaching an almost stalling configuration. These are characterised by the largest $r_{\rm out}/r_e$ values, such as e.g. Maisie ($z=11.44$) and other galaxies (blue circles).    

Finally, as expected in our scenario, Red Monsters are still in the dust accumulation phase or have just entered the super-Eddington one. Hence their outflows are still absent or have limited size (e.g., CAPERS\_UDS\_z11). 

\subsubsection{Metallicity}\label{subsec:metallicity}
The last quantity we scrutinise is the metallicity. {Where observational metallicity estimates are available, the agreement
with the AFM prediction is object-dependent. A quantitative comparison,
including the residuals and their formal significance, is given in
Appendix~\ref{app:residuals}.
}

{The AFM predicts significant metallicities for Blue Monsters} (median $\log (Z/Z_\odot) = -1.17$), although there is a scatter of about $0.2$ dex around that value. Galaxies at $z<11$ lie close to or above the median, while those at higher redshifts are instead well below such value. This indicates a redshift trend, which is generally expected. 


{Within the evolutionary scenario proposed here,
a Blue Monster may emerge from a previous Red Monster
phase, during which metals and dust accumulate.
The duration of this phase depends on the star-formation
history and cannot be inferred directly from the
instantaneous mass-assembly timescale
$M_\star/{\rm SFR}$. Red Monsters residing in massive
halos may retain a substantial fraction of the metals
produced by their stellar populations, potentially
leading to relatively high metallicities.}

{The comparison with the available observational estimates is, however, not
uniform across the sample. For EGS-25297, AFM predicts
$\log(Z/Z_\odot)=-0.65$, compared with the SED-based estimate
$\log(Z/Z_\odot)=-1.00\pm0.08$ from \citet{Donnan26}. The $0.35$ dex
offset is significant relative to the quoted statistical uncertainty
($\simeq4.4\sigma$), and therefore represents a tension between the model
prediction and the SED-based estimate. An even larger absolute discrepancy
is found for CEERS2-588, for which AFM predicts
$\log(Z/Z_\odot)=-1.04$, compared with the observed
$-0.14^{+0.10}_{-0.13}$, corresponding to an offset of about $-0.90$ dex.}

{Conversely, several other objects show substantially better agreement
(Appendix~\ref{app:residuals}). We therefore conclude that, while the AFM
predicts the overall metallicity scale of the sample without a significant
systematic offset, the present object-by-object comparison shows substantial
scatter and a few significant discrepancies.}

\subsection{Evolutionary scenario} \label{sec:evol}
The question that naturally arises from the above analysis of the predicted super-early galaxy properties is whether their apparent diversity can be reconciled with a common evolutionary path. In other words, can we identify the leading physical processes that give rise to blue (and red) monsters?

To answer this question, we take advantage of the detailed study \citep{Ferrara24b} of a representative Blue Monster, GS-z14-0 at $z=14.18$, whose observed and predicted properties are listed in Tab. \ref{tab:observed_props} and Tab. \ref{tab:predicted_props}, respectively. In that work the observed properties and spectrum of GS-z14-0 were obtained using a physical model of its star formation history (SFH), which we use here as a template to propose a possible common evolutionary path of super-early galaxies. A schematic view of the SFH as a function of lookback time, $t_L$, from the observed epoch is presented in Fig. \ref{fig:evo_scheme}. Its main features are summarized below.

Star formation activity is largely suppressed by molecular cooling dissociation in \quotes{minihalos}, i.e. halos with virial temperature $T_{\rm vir} \le 10^4$ K (see e.g. \citealt{Ciardi05}). Hence, stars begin to form at $z = z_*$, corresponding to $\simeq 170\ \rm Myr$, when growing halos cross that boundary ($M_0 \simeq 10^8\ M_\odot$) and Ly$\alpha$ cooling becomes efficient. The halo mass growth with redshift can be obtained by time-integrating the cosmological accretion of matter. A handy fit to $M(z)$ is provided by \citet{Correa15}:
\be
M(z) = M_0 \left(\frac{1+z}{1+z_0}\right)^{a_0} e^{a_1 (z-z_0)},
\label{eq:Mhalo}
\ee
where $a_0$ and $a_1$ are introduced in eq. \ref{eq:a}. 

Following a very bursty and intermittent initial phase\footnote{{This internal AFM bursty phase should not be confused
with models in which stochastic SFR fluctuations up-scatter more numerous
low-mass systems into the bright UV population
\citep[e.g.][]{Sun23,Pallottini23,Mason23,Gelli25}. Such stochasticity
would broaden the deterministic ${\rm SFR}(M,z)$ mapping adopted here.
The scenarios can be distinguished through spatially extended/offset dust
\citep{Carniani25,Schouws24}, resolved metal-line outflows
\citep{Carniani26}, enriched circumgalactic gas \citep{Higginson26}, and
independent SFH constraints. {Additional discriminants between these scenarios include UV variability and AGN-sensitive spectroscopic diagnostics, such as high-ionization emission lines and broad permitted-line components.}
}}, star formation becomes more continuous as the galaxy
grows in stellar mass.
As a result, dust produced by SNe continues to accumulate, increasingly attenuating the galaxy's light, eventually completely obscuring the galaxy. As the SFR continues to increase, sustained by vigorous mass accretion (see eq. \ref{eq:a}), the increased dust/metal content lowers the value of the critical specific star formation, sSFR$^*$, and the galaxy becomes super-Eddington and launches a powerful outflow.

The outflow very rapidly (on a timescale of a few Myr) displaces the dust and gas to a radius $r_{\rm out} > r_e$, as we have seen in the previous Section. The outflow also quenches the SFR due to the decreased availability of gas. At that point, the sSFR falls below sSFR$^*$ and eventually the outflow radius approaches its observed value. This is found (Sec. \ref{subsec:outflow}) to be on average $r_{\rm out}/r_e \approx 7$, which implies outflows extending to $\approx$ kpc scales.

This evolutionary path is schematically illustrated by Fig. \ref{fig:evo_scheme}. Also shown, based on the analysis performed in Sec. \ref{sec:results} above, are the locations of EGS-z11-R0 (left panel) and GS-z14-0 (right). The comparison highlights the different evolutionary paths and phases of Red vs. Blue Monsters. According to AFM, the $z>10$ Red Monsters are observed either in the dust accumulation phase or immediately after they become super-Eddington, when the outflow starts to clear the dust cocoon. In between, we predict the existence of a possibly brief phase of complete obscuration in which the galaxies are too faint and red to be seen by JWST (i.e., JWST-dark galaxies), with the possible exception of some MIRI bands. A detailed modeling of this phase (SED, number counts) is deferred to further study.

Blue Monsters first appear at the end of the rapid quenching phase, when the outflow is still ongoing but gradually running out of power. Observationally, evidence of quenching in CEERS2-588 from MIRI spectra has been reported by \citet{Harikane26}, in line with the present results. They remain blue and bright even when the outflow stalls at $r_{\rm out}$ and the galaxy settles into a more moderate SFR. The proposed evolution also explains why extended Blue Monsters show almost featureless spectra and weak emission lines, consistent with the post-starburst scenario we propose here.

Interestingly, the above evolutionary scenario is strongly supported by the recent study by \citet{Rojas26} who studied $z\approx 8$ Blue Monsters found in the BoRG-JWST survey. These authors conclude that these galaxies are seen in a fading phase after a first or second major burst occurred at $z>12$. 

In principle, a blue phase can also occur at large lookback times, i.e., earlier during the evolution and at the beginning of the dust accumulation phase. Although the SFR -- and hence the UV luminosity -- is probably too low for the galaxy to be detected, this evolutionary phase might be within reach of lensing experiments, such as the GLIMPSE survey \citep{Atek25}. It would be interesting in the future to use the AFM to make detailed predictions concerning this phase.

To further corroborate the above evolutionary scenario, in Fig. \ref{fig:z_vs_vc} we plot the evolutionary tracks (green lines) of galaxies starting to form stars at different redshifts in the range $9 < z_* < 25$. The tracks are shown in the redshift–circular velocity plane (proxy for halo mass); the stellar mass isocontours (blue dotted) allow us to infer $M_*$ along the track. In both panels, we show for reference the curves for $A_V = 0.5$ (our working definition of a red/obscuRed Monster) and $A_V = 1.5$. These curves refer to pre-outflow values; as the outflow displaces the dust, $A_V$ decreases to a much lower value, as clarified when discussing Fig. \ref{fig:properties}. The location of the red curve is set by the Eddington condition sSFR = sSFR$^*$ (Eq. \ref{eq:SEdd_condition}). We note that both the $A_V$ and the critical sSFR curves depend on the galaxy properties, such as $\lambda$ and $\Delta_{t_*}$, and therefore vary from galaxy to galaxy.

To illustrate our point, in Fig. \ref{fig:z_vs_vc} we compare a Red Monster (EGS-z11-R0) with a Blue Monster (GS-z14-0). As expected, EGS-z11-R0, with its inferred $v_c=156.87\ \kms$, is located well into the obscured region ($A_V=0.9$), into which it entered at $z = 13.5$, having started to form stars at $z_* \simeq 20$. Our model predicts that EGS-z11-R0 will continue to accumulate dust and become severely attenuated for a long time (up to $z\simeq 7.5$), when it will have reached $v_c = 320\ \kms$ corresponding to a halo mass $\log M = 11.95$. At that point, the outflow starts clearing the dust. Strikingly, this halo mass is comparable to that inferred for the other $z=9.94$ Red Monster EGS-25297, which is in the super-Eddington phase. We caution, however, that if the accumulation phase lasts for a long time, allowing the galaxy to become very large, our assumption of a central, compact source driving the outflow might no longer be satisfied. A firm conclusion on this issue requires further work.

{We also notice that EGS-z11-R0 has only an upper limit on its effective
radius, $r_e<280$ pc, which does not by itself establish that the
source belongs to the extended ($r_e\gtrsim150$ pc) population.
The AFM solution discussed here is therefore conditional on adopting
the quoted upper limit as its characteristic size. A more compact
configuration, potentially including a significant AGN contribution,
cannot presently be excluded.}

{In fact, the detection of high-ionization lines in
EGS-z11-R0 leaves open the possibility of an AGN contribution to its
UV luminosity. If significant, such a contribution would reduce the
stellar SFR associated with the observed UV emission and consequently
modify some of the AFM-inferred quantities. The present solution
should therefore be interpreted as the stellar-dominated limiting case.}

The situation is dramatically different for GS-z14-0 ($z=14.18$), which is instead actively powering an outflow whose current outer radius is $r_{\rm out} \approx 5\times r_e = 1.3\ \rm kpc$. This system started to form stars at $z_* \simeq 24$ and passed through a relatively brief obscured phase lasting from $z=19.2$ to $16.6$, or 45 Myr. As the galaxy's sSFR $\approx 2 \times \rm sSFR^*$ (Table \ref{tab:predicted_props}), the outflow is still ongoing, sustained by a significant SFR of $19\ M_\odot\ \rm yr^{-1}$. Interestingly, \citet[]{Carniani26} report evidence for CIII]$\lambda\lambda$1907,1909 line emission from the outflowing material at least up to a radius $\simgt 400$ pc. From the spectra they also set an upper limit to the star formation efficiency $\epsilon_* < 0.08$, which aligns well with the value found here ($\epsilon_* = 0.0273$).

In summary, the above scenario, highlighting the key role played by radiation-driven outflows in the early universe, provides a coherent interpretation framework to understand the diversity of super-early galaxies and their evolutionary properties, while making testable and specific predictions for ongoing high-redshift galaxy searches.

\subsection{Possible lower-redshift imprints}
\label{subsec:lowz}
{
Although we have focused on $z>10$, the AFM mechanism is not restricted
to this epoch. The super-Eddington condition ${\rm sSFR}>{\rm sSFR}^*$
can also be satisfied at later times, although the declining cosmic sSFR
should make radiation-driven dust-clearing episodes progressively less
common. Possible lower-redshift counterparts include obscured or JWST-dark
systems followed by high-sSFR, weakly attenuated phases
\citep[e.g.][]{Madau14,Barrufet23,Sun26}; their relative abundance provides
an integral constraint on the duration of the obscured phase.
}

{
More direct tests follow from the predicted spatial redistribution of
dust and metals. Radiation-driven clearing should produce dust emission
extended or displaced relative to the rest-UV stellar component, as
observed at $z\simeq4$--$6$ \citep{Mitsuhashi24}, together with enriched
circumgalactic gas traced by extended [CII]/dust emission
\citep[e.g.][]{Fujimoto19,Ginolfi20} or metal absorption
\citep{Zhu26}. These provide independent tests of the Red-to-Blue
evolutionary picture beyond $z>10$.
}

\section{Compact galaxies subset} \label{sec:compact}
So far we have concentrated on the 17 galaxies selected for the detailed
AFM analysis. Of the remaining sources with size constraints, 11 have
measured or upper-limit values $r_e\lesssim150$ pc and constitute the
compact subset. Explaining such extremely compact structure is very challenging within the framework of our model. A practical example involving the prototypical compact galaxy GHZ2 ($r_e = 105 \pm 9$ pc) clarifies the difficulty.

The very small size of GHZ2 requires a slowly rotating halo, with $\lambda = 0.0097$ -- a value corresponding to a $2.3\sigma$ deviation from the mean. In addition, the highly compact structure together with the large stellar mass would make the galaxy extremely obscured (see eq. \ref{eq:tauUV}). To decrease $A_V$ to the observed value of $\sim 0.04$, the outflow radius would need to be $r_{\rm out} \approx 23.6$ kpc, i.e., 2.5 times larger than the virial radius of GHZ2 (9.5 kpc). However, according to eq. \ref{eq:SEdd_condition}, GHZ2 should not have reached the outflow phase so far. This contradiction suggests that compact galaxies have a different nature from the \quotes{normal}, extended population.

Taken together with the presence of strong high-ionization emission lines, and in particular NIV]$\lambda$1486 \citep[]{Harikane26}, a more plausible explanation is that the emission from compact super-early galaxies is instead dominated by an AGN.  Indeed, AGN activity in compact galaxies such as GN‑z11 and GHZ2, based on the detection of high‑ionization emission lines, has been reported \citep[]{Maiolino24b, Castellano24}. In this case, the galaxy rest‑frame UV continuum must be dominated by emission from an AGN accretion disk. Importantly, the presence of the AGN in compact systems would strongly affect the properties inferred from the observations, thus making them rather uncertain.

To quantify the above statement, we proceed as follows. Let us suppose that compact super-early galaxies host a central AGN whose UV luminosity $L_{\rm AGN} = x L_{\rm UV}$, where $L_{\rm UV}$ is the galaxy luminosity; further assume that the galaxy light follows a Sercic profile with $n=1$, i.e. exponential, with length scale $r_s=r_e/1.678$. Then the fraction of the host galaxy light outside a radius $r$ is 
\be
f_{\rm out}(r) = e^{-r/r_s}\left(1+\frac{r}{r_s}\right)
\ee
If the PSF dominates within a radius $r_0 \approx 1$ FWHM, the host luminosity outside that radius is $L_{\rm UV, out} = L_{\rm UV} f_{\rm out}(r_0)$. Hence, the galaxy \textit{cannot} be detected if
\be
\frac{L_{\rm UV, out}}{L_{\rm AGN}} = \frac{1}{x} f_{\rm out}(r_0) < \epsilon_{\rm PSF}, 
\ee
where $\epsilon_{\rm PSF} \approx 0.05$ is the fractional PSF-subtraction floor at $r_0$. Quantitatively, we fix $r_e = 400$ pc, an average value for the extended-galaxy subset, and hence $r_s=238$ pc, or 0.06" at $z=11$. To estimate $r_0$, we note that for JWST/NIRCam the PSF FWHM is approximately diffraction-limited, FWHM$\sim \lambda/D$, and for F444W it is of order 0.15". From the above expressions, we then find $x =5.7$, which implies in terms of the host fraction of the total light
\be
\frac{L_{\rm UV}}{L_{\rm tot}} = \frac{1}{1+x} = 0.15.
\ee
{Under the assumptions above, an AGN contribution exceeding approximately
85\% of the total UV luminosity would be sufficient to render the extended
host undetectable at the adopted PSF-subtraction floor, causing the source
to appear point-like. We stress that this value is a model-dependent
detectability threshold, rather than a measurement of the AGN fraction
in the compact sources. An AGN-dominated UV continuum therefore provides
one possible explanation for the compact-galaxy subset.}

\section{Summary} \label{sec:summary}

In this work, we have investigated the physical origin of the diversity observed among spectroscopically confirmed super-early galaxies at $z>10$ within the framework of the Attenuation-Free Model (AFM). {Starting from 32 spectroscopically confirmed galaxies, we apply the
detailed AFM analysis to the 17 systems with suitable size constraints,
deriving their halo masses, star formation efficiencies, metallicities,
and outflow properties.} {Based on these results, the AFM consistently explains the coexistence of extremely blue and significantly reddened systems at cosmic dawn.}

Our analysis supports an evolutionary scenario in which super-early galaxies undergo a sequence of phases regulated by the interplay between star formation, dust production, and radiation-driven feedback. Galaxies initially build up dust through supernova enrichment, entering an obscured ``Red Monster'' phase characterized by significant attenuation and moderate-to-high metallicity. As the star formation rate increases, galaxies can reach a super-Eddington regime, triggering powerful radiation-driven outflows that displace dust and gas to kiloparsec scales. This process rapidly reduces the effective optical depth, giving rise to the observed population of bright, UV-transparent ``Blue Monsters''. Within this framework, sources such as EGS-z11-R0 are naturally interpreted as systems caught during the dust-enshrouded phase preceding the emergence of blue galaxies.

Overall, the AFM provides a {self-consistent} physical picture that connects the observed properties of super-early galaxies to their underlying evolutionary paths. It highlights the central role of radiation pressure and outflows in shaping early galaxy evolution, and makes specific, testable predictions regarding dust distribution, metallicity, and the existence of heavily obscured, potentially JWST-dark progenitors. These results have important implications for our understanding of early star formation, chemical enrichment, and the buildup of the first luminous structures in the Universe.

\begin{itemize}
\item[\color{red}$\blacksquare$] {The AFM self-consistently infers a broad set of physical properties of
$z>10$ galaxies, including independently testable quantities such as
metallicity, while $M_{\rm UV}$ and $\beta$ provide input-dependent
consistency checks.}

\item[\color{red}$\blacksquare$] Super-early galaxies exhibit moderate star formation efficiencies ($0.01 \lesssim \epsilon_* \lesssim 0.05$) and reside in massive halos ($\log M/M_\odot \sim 10.7$), corresponding to rare ($\nu_\delta\simeq4.3$--$6.8$, median $\simeq5.2$) peaks of the density field.

\item[\color{red}$\blacksquare$] Radiation-driven outflows are a ubiquitous and key ingredient of early galaxy evolution, displacing dust to radii $\sim 5$--$10\,r_e$ and enabling the transition from dust-obscured to UV-bright phases.

\item[\color{red}$\blacksquare$] {Within the AFM framework, the coexistence of
extremely blue and significantly reddened systems
can be consistently interpreted as different stages
of a possible Red-to-Blue evolutionary sequence,
although alternative explanations are not excluded.}

\item[\color{red}$\blacksquare$] {Compact ($r_e\lesssim150$ pc) systems
are difficult to reconcile within the AFM framework. An AGN-dominated
UV continuum provides a possible explanation; our simple PSF argument
shows that an AGN contribution $\gtrsim85\%$ would be sufficient to hide
an otherwise extended host under the assumptions adopted here.}
\end{itemize}
Future progress will crucially rely on larger samples and deeper observations with {\it JWST} and ALMA, which will be essential to test these predictions, constrain the dust and gas distribution in super-early galaxies, and clarify the physical processes driving their rapid evolution at cosmic dawn.

\section*{Acknowledgments}
We thank H. Atek, T. Bakx, L. Bisigello, M. Castellano, R. Ellis, C. Kobayashi, M. Llerena,  R. Maiolino, D. McLeod, I. Mitsuhashi,  T. Morishita, Y. Nakazato, L. Napolitano, L. Pentericci, P. P\'erez-Gonz\'alez, G. Roberts-Borsani, P. Santini, R. Somerville, M. Stiavelli, S. Tacchella, M. Tang, and J. Zavala  for valuable discussions and/or sharing galaxy data. A special thank from AF (he knows why) to M. Ouchi. This work is supported by the ERC Advanced Grant INTERSTELLAR H2020/740120 (PI: Ferrara), and in part by grant NSF PHY-2309135 to the Kavli Institute for Theoretical Physics. This work was started while AF was a Distinguished Visitor at the Institute of Astronomy, Cambridge University, whose generous support and hospitality are kindly acknowledged. 

\bibliographystyle{plainnat}
\bibliography{paper}

\appendix

\section{Residuals between model and observations}
\label{app:residuals}
{Here we quantify the residuals between the AFM-derived quantities
and the corresponding observed values for the 17 galaxies included
in the detailed analysis (Table~\ref{tab:residuals}). We define
$\Delta X \equiv X_{\rm AFM}-X_{\rm obs}$.
For metallicity, the AFM values are converted to $\log(Z/Z_\odot)$ before computing the residuals. 
}

{The residuals show no evidence for a significant systematic offset between
the AFM-derived and observed quantities, with mean differences of
$\langle\Delta M_{\rm UV}\rangle=0.05$ mag,
$\langle\Delta\beta\rangle=-0.09$, and
$\langle\Delta\log(Z/Z_\odot)\rangle=-0.003$ dex.
The corresponding rms scatter is, however, appreciable: $0.79$ mag,
$0.37$, and $0.39$ dex, respectively.
Using the quoted observational statistical uncertainties alone, eight
galaxies show formally $>2\sigma$ residuals in $\beta$, with the largest
discrepancy found for CEERS2-7929 ($\simeq6.4\sigma$).
For metallicity, three objects formally exceed $2\sigma$: EGS-25297,
CEERS2-588, and GS-z14-0. The largest absolute metallicity discrepancy
is found for CEERS2-588 ($\Delta\log Z=-0.90$ dex), whereas the high
formal significance for GS-z14-0 results primarily from its very small
quoted statistical uncertainty.
We stress that these significances do not include systematic uncertainties
in the observationally inferred quantities or uncertainties propagated
through the AFM relations, and therefore should not be interpreted as a
goodness-of-fit test.}

\begin{table}[ht!]
\centering
\caption{
{Residuals between the AFM-derived quantities and the corresponding observed
values for the 17 galaxies included in the detailed analysis.
We define $\Delta X\equiv X_{\rm AFM}-X_{\rm obs}$.
For metallicity, the AFM values listed in
Table~\ref{tab:predicted_props} as $Z/Z_\odot$ have been converted to
$\log(Z/Z_\odot)$ before computing the residual.
A superscript ``$*$'' marks residuals whose formal significance exceeds
$2\sigma$ when only the quoted observational statistical uncertainty is
considered. These significances should not be interpreted as a full
goodness-of-fit statistic, since they do not include systematic uncertainties
in the observational inference or uncertainties propagated through the AFM
relations.
The rms residuals are $0.79$ mag, $0.37$, and $0.39$ dex for
$M_{\rm UV}$, $\beta$, and $\log(Z/Z_\odot)$, respectively, while the
corresponding mean residuals are $0.05$ mag, $-0.09$, and $-0.003$ dex.}
}
\label{tab:residuals}
\vskip 1truecm
\begin{tabular}{lccc}
\hline
\hline
Galaxy &
$\Delta M_{\rm UV}$ [mag] &
$\Delta\beta$ &
$\Delta\log(Z/Z_\odot)$ [dex] \\
\hline
EGS-25297      & $-1.859$        & $+0.304^{*}$ & $+0.348^{*}$ \\
UNCOVER-26185  & $+0.236^{*}$    & $+0.040$     & $+0.057$     \\
CEERS2-7929    & $-0.789$        & $-0.637^{*}$ & ---          \\
CAPERS UDS z10 & $-0.375^{*}$    & $+0.180$     & ---          \\
EGS-22637      & $+0.621^{*}$    & $+0.040$     & $+0.537$     \\
MoM-z11-2      & $-0.441$        & $-0.752^{*}$ & $-0.519$     \\
MoM-z11-1      & $+0.403$        & $+0.127$     & $+0.405$     \\
CAPERS UDS z11 & $-0.762^{*}$    & $+0.243^{*}$ & ---          \\
CEERS2-588     & $-0.096$        & $-0.626^{*}$ & $-0.899^{*}$ \\
MAISIE         & $+1.177^{*}$    & $+0.032$     & ---          \\
EGS-z11-R0     & $+1.597^{*}$    & $-0.555$     & ---          \\
UNCOVER-z12    & $+0.328$        & $+0.076$     & $+0.048$     \\
GS-z12-0       & $-0.449^{*}$    & $-0.425^{*}$ & $+0.113$     \\
UNCOVER-z13    & $+0.717$        & $+0.108$     & $+0.174$     \\
PAN-z14-1      & $+0.534^{*}$    & $-0.193^{*}$ & $+0.197$     \\
GS-z14-1       & $+0.230$        & $+0.460^{*}$ & $-0.250$     \\
GS-z14-0       & $-0.153$        & $+0.110$     & $-0.242^{*}$ \\
\hline
Mean & $+0.054$ & $-0.086$ & $-0.003$ \\
RMS  & $0.794$  & $0.370$  & $0.394$  \\
\hline
\end{tabular}
\end{table}

\end{document}